\documentclass[%
 reprint,
 superscriptaddress,
 amsmath,amssymb,
 nofootinbib,
nobibnotes,
 aps,
 prd,
]{revtex4-2}

\preprint{Accepted to PRD}

\usepackage{graphicx}
\usepackage{dcolumn}
\usepackage{bm}
\usepackage{siunitx}
\usepackage{multirow}
\usepackage{soul}
\usepackage{lineno}
\usepackage{url}
\usepackage{hyperref}
\hypersetup{colorlinks=true,linkcolor=blue, linktocpage}

\begin{document}

\preprint{YOLO}

\title{Transforming a rare event search into a not-so-rare event search\\in real-time with deep learning-based object detection}

\author{J.~Schueler}
\thanks{jschueler1@unm.edu}
\affiliation{Department of Physics and Astronomy, University of New Mexico, Albuquerque, NM, 87131, USA}

\author{H.~M.~Ara\'ujo}
\affiliation{Department of Physics, Blackett Laboratory, Imperial College London, London, SW7 2AZ, UK}

\author{S.~N.~Balashov}
\affiliation{Particle Physics Department, STFC Rutherford Appleton Laboratory, Didcot, OX11 0QX, UK}

\author{J.~E.~Borg}
\affiliation{Luleå University of Technology, 97187 Luleå, Sweden}

\author{C.~Brew}
\affiliation{Particle Physics Department, STFC Rutherford Appleton Laboratory, Didcot, OX11 0QX, UK}

\author{F.~M.~Brunbauer}
\affiliation{CERN, 1211 Geneva 23, Switzerland}

\author{C.~Cazzaniga}
\affiliation{ISIS Neutron and Muon Source, STFC Rutherford Appleton Laboratory, Didcot, OX11 0QX, UK}

\author{A.~Cottle}
\affiliation{Department of Physics, Keble Road, University of Oxford, Oxford, OX1 3RH, UK}

\author{C.~D.~Frost}
\affiliation{ISIS Neutron and Muon Source, STFC Rutherford Appleton Laboratory, Didcot, OX11 0QX, UK}

\author{F.~Garcia}
\affiliation{Helsinki Institute of Physics, University of Helsinki, FI-00014 Helsinki, Finland}

\author{D.~Hunt}
\affiliation{Department of Physics, Keble Road, University of Oxford, Oxford, OX1 3RH, UK}

\author{A.~C.~Kaboth}
\affiliation{Department of Physics, Royal Holloway University of London, Egham, TW20 0EX, UK}

\author{M.~Kastriotou}
\affiliation{ISIS Neutron and Muon Source, STFC Rutherford Appleton Laboratory, Didcot, OX11 0QX, UK}

\author{I.~Katsioulas}
\affiliation{School of Physics and Astronomy, University of Birmingham, Birmingham, B15 2TT, UK}

\author{A.~Khazov}
\affiliation{Particle Physics Department, STFC Rutherford Appleton Laboratory, Didcot, OX11 0QX, UK}

\author{P.~Knights}
\affiliation{School of Physics and Astronomy, University of Birmingham, Birmingham, B15 2TT, UK}

\author{H.~Kraus}
\affiliation{Department of Physics, Keble Road, University of Oxford, Oxford, OX1 3RH, UK}

\author{V.~A.~Kudryavtsev}
\affiliation{Department of Physics and Astronomy, University of Sheffield, Sheffield, S3 7RH, UK}

\author{S.~Lilley}
\affiliation{ISIS Neutron and Muon Source, STFC Rutherford Appleton Laboratory, Didcot, OX11 0QX, UK}

\author{A.~Lindote}
\affiliation{LIP -- Laborat\'{o}rio de Instrumenta\c{c}\~{a}o e F\'{\i}sica Experimental de Part\'{\i}culas, University of Coimbra, P-3004-516 Coimbra, Portugal}

\author{M.~Lisowska}
\let\comma,
\affiliation{CERN, 1211 Geneva 23, Switzerland}
\affiliation{Universit\'e Paris Saclay, F-91191 Gif-sur-Yvette, France}

\author{D.~Loomba}
\affiliation{Department of Physics and Astronomy, University of New Mexico, Albuquerque, NM, 87131, USA}

\author{M.~I.~Lopes}
\affiliation{LIP -- Laborat\'{o}rio de Instrumenta\c{c}\~{a}o e F\'{\i}sica Experimental de Part\'{\i}culas, University of Coimbra, P-3004-516 Coimbra, Portugal}

\author{E.~Lopez~Asamar}
\affiliation{Departamento de Fisica Teorica, Universidad Autonoma de Madrid, 28049 Madrid, Spain}

\author{P.~Luna~Dapica}
\affiliation{ISIS Neutron and Muon Source, STFC Rutherford Appleton Laboratory, Didcot, OX11 0QX, UK}

\author{P.~A.~Majewski}
\affiliation{Particle Physics Department, STFC Rutherford Appleton Laboratory, Didcot, OX11 0QX, UK}
\affiliation{Department of Physics, Blackett Laboratory, Imperial College London, London, SW7 2AZ, UK}

\author{T.~Marley}
\affiliation{Department of Physics, Blackett Laboratory, Imperial College London, London, SW7 2AZ, UK}

\author{C.~McCabe}
\affiliation{Department of Physics, King’s College London, London, WC2R 2LS, UK}

\author{L.~Millins}
\affiliation{School of Physics and Astronomy, University of Birmingham, Birmingham, B15 2TT, UK}
\affiliation{Particle Physics Department, STFC Rutherford Appleton Laboratory, Didcot, OX11 0QX, UK}

\author{A.~F.~Mills}
\affiliation{Department of Physics and Astronomy, University of New Mexico, Albuquerque, NM, 87131, USA}

\author{M.~Nakhostin}
\affiliation{Department of Physics, Blackett Laboratory, Imperial College London, London, SW7 2AZ, UK}
\affiliation{Particle Physics Department, STFC Rutherford Appleton Laboratory, Didcot, OX11 0QX, UK}

\author{R.~Nandakumar}
\affiliation{Particle Physics Department, STFC Rutherford Appleton Laboratory, Didcot, OX11 0QX, UK}

\author{T.~Neep}
\affiliation{School of Physics and Astronomy, University of Birmingham, Birmingham, B15 2TT, UK}

\author{F.~Neves}
\affiliation{LIP -- Laborat\'{o}rio de Instrumenta\c{c}\~{a}o e F\'{\i}sica Experimental de Part\'{\i}culas, University of Coimbra, P-3004-516 Coimbra, Portugal}

\author{K.~Nikolopoulos}
\affiliation{School of Physics and Astronomy, University of Birmingham, Birmingham, B15 2TT, UK}
\affiliation{University of Hamburg, 22767, Hamburg, Germany}

\author{E.~Oliveri}
\affiliation{CERN, 1211 Geneva 23, Switzerland}

\author{L.~Ropelewski}
\affiliation{CERN, 1211 Geneva 23, Switzerland}

\author{V.~N.~Solovov}
\affiliation{LIP -- Laborat\'{o}rio de Instrumenta\c{c}\~{a}o e F\'{\i}sica Experimental de Part\'{\i}culas, University of Coimbra, P-3004-516 Coimbra, Portugal}

\author{T.~J.~Sumner}
\affiliation{Department of Physics, Blackett Laboratory, Imperial College London, London, SW7 2AZ, UK}

\author{J.~Tarrant}
\affiliation{Technology Department, STFC Rutherford Appleton Laboratory, Didcot, OX11 0QX, UK}

\author{E.~Tilly}
\affiliation{Department of Physics and Astronomy, University of New Mexico, Albuquerque, NM, 87131, USA}

\author{R.~Turnley}
\affiliation{ISIS Neutron and Muon Source, STFC Rutherford Appleton Laboratory, Didcot, OX11 0QX, UK}

\author{R.~Veenhof$\,$}
\affiliation{CERN, 1211 Geneva 23, Switzerland}

\collaboration{MIGDAL Collaboration}

\begin{abstract}
Deep learning-based object detection algorithms enable the simultaneous classification and localization of any number of objects in image data. Many of these algorithms are capable of operating in real-time on high resolution images, attributing to their widespread usage across many fields. We present an end-to-end object detection pipeline designed for rare event searches for the Migdal effect, at real-time speeds, using high-resolution image data from the scientific CMOS camera readout of the MIGDAL experiment. The Migdal effect in nuclear scattering, critical for sub-GeV dark matter searches, has yet to be experimentally confirmed, making its detection a primary goal of the MIGDAL experiment. The Migdal effect forms a composite rare event signal topology consisting of an electronic and nuclear recoil sharing the same vertex. Crucially, both recoil species are commonly observed in isolation in the MIGDAL experiment, enabling us to train YOLOv8, a state-of-the-art object detection algorithm, on real data. Topologies indicative of the Migdal effect can then be identified in science data via pairs of neighboring or overlapping electron and nuclear recoils. Applying selections to real data that retain $99.7\%$ signal acceptance in simulations, we demonstrate our pipeline to reduce a sample of 20 million recorded images to fewer than 1,000 frames, thereby transforming a rare search into a much more manageable search. More broadly, we discuss the applicability of using object detection to enable data-driven machine learning training for other rare event search applications such as neutrinoless double beta decay searches and experiments imaging exotic nuclear decays.

\end{abstract}


\maketitle


\section{\label{sec:intro}Introduction}

\begin{figure*}[htbp]
\centering
\includegraphics[width=\textwidth]{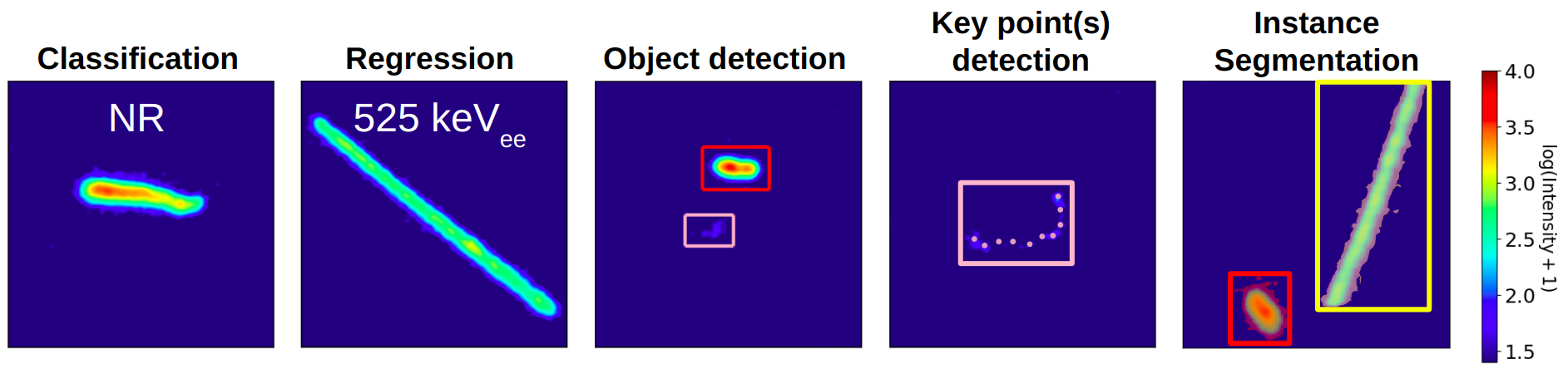}
\caption{Example outputs of common CNN-based computer vision tasks illustrated on snippets of image data recorded in the MIGDAL detector. Referring to each frame from left to right: (i) Classification maps an image input to a discrete set of outputs. The example shown here is for an image classifier trained for particle identification, where the classifier predicts a nuclear recoil (NR) from neutron scattering in the input image. (ii) Regression maps the input to a continuous set of outputs. The example illustrated here is for a model trained to reconstruct energies, so the regression model reconstructs the energy present in the input image  as \SI{525}{keV} of visible energy. (iii) Object detection algorithms simultaneously classify and localize any number of objects in a single image. The two bounding boxes shown were predicted by our trained YOLOv8 pipeline (see Sec.$\,$\ref{sec:YOLOpipeline}) and indicate that the algorithm detected a NR track (red box) and an ER track (pink box). (iv) Key point detection takes object detection a step further and identifies key points within bounding boxes; shown here is an example of particle trajectory fitting with key points. (v) Finally, within each classified bounding box, instance segmentation assigns every pixel as belonging to the object-class or not. This example shows translucent segmentation masks overlaid on the NR and proton within the red and yellow bounding boxes, respectively, designating the pixels that the algorithm assigned as belonging to the track. \label{fig:CV}}
\end{figure*}

Convolutional neural networks (CNNs) as backbones for computer vision systems have found remarkable success in extracting meaningful information from image and video data. AlexNet$\,$\cite{NIPS2012c399862d} was one of the first major breakthroughs in CNN-based computer vision, where it achieved a Top-5 image classification error rate that was more than 10 percentage points lower than its closest competition in the ImageNet$\,$\cite{5206848} 2012 contest. This result brought deep learning and CNNs to the forefront of modern computer vision research. Since then, CNNs have enabled a host of other computer vision applications including regression predictions of image inputs, object detection, key point detection, and instance segmentation as exemplified with the image data from our experiment shown in Fig.$\,$\ref{fig:CV}.

Image classification and regression are among the simplest computer vision applications, where images are passed through an algorithm (often a CNN) and mapped to discrete and continuous sets of outputs, respectively. Common examples of classification and regression tasks in high energy physics are particle identification$\,$\cite{Aurisano_2016} and energy reconstruction$\,$\cite{Baldi:2018qhe}, both of which can be performed simultaneously with suitable choice of loss function and model architecture$\,$\cite{Belayneh:2019vyx}. 

Object detection is more complicated and involves the simultaneous classification and localization of any number of objects in an input image. The output of an object detection algorithm consists of bounding boxes surrounding each identified object (center panel of Fig.$\,$\ref{fig:CV}), where each bounding box has an associated classification prediction. The Regions with CNN features (R-CNN) algorithm$\,$\cite{girshick2014rich} demonstrated the first usage of CNNs for object detection in 2014, and since then, many refinements and new approaches to deep learning-based object detection have been introduced$\,$\cite{fasterrcnn,liu2016ssd,lin2017focal,carion2020end}. Due to their balance of speed and accuracy, the You Only Look Once (YOLO)$\,$\cite{redmon2016you} family of algorithms are among the most popular object detection algorithms for real-time applications. YOLO has been continuously improved upon (see Ref.$\,$\cite{terven2023comprehensive} for a comprehensive history) with YOLOv8$\,$\cite{yolov8}, the version used in this work, being among the current state of the art in fast object detection.

Object detection is used widely in applications spanning many fields$\,$\cite{GHASEMI2022103661, 9445365, yangmed, 2018ISPAr42.3.1915W, 9352144}, and is beginning to see more usage in the physical sciences. In astronomy, YOLO in particular has found success in galaxy detection and identification$\,$\cite{GONZALEZ2018103}, and the recently developed YOLO-CIANNA$\,$\cite{2024arXiv240205925C} outperformed the winner of the Square Kilometre Array Science Data Challenge 1$\,$\cite{Bonaldi:2020ukl}, demonstrating YOLO's efficacy in analyzing large astronomical datasets. Object detection and semantic segmentation -- the class-assignment of individual pixels in an image -- have garnered recent interest in neutrino physics, with MicroBooNE$\,$\cite{MicroBooNE:2016pwy,MicroBooNE:2016dpb,MicroBooNE:2020yze,MicroBooNE:2020hho,MicroBooNE:2022tdj} using semantic segmentation for track reconstruction in their search for the anomalous low energy excess observed by MiniBooNE$\,$\cite{MicroBooNE:2021pvo,PhysRevD.103.052002}. YOLO has also been proposed as a way to improve long distance supernova burst trigger performance in DUNE$\,$\cite{osti_1873684}. Other novel object detection approaches have also been proposed for particle physics applications outside of neutrino experiments$\,$\cite{Kieseler:2020wcq}. 

Machine learning (ML) applications in particle physics commonly fall under the umbrella of supervised learning, where models are trained on data that is explicitly labeled. The form of the labels associated with data will depend on the task at hand.\footnote{The outputs in the image snippets of Fig.$\,$\ref{fig:CV} exemplify the form of labels for the five shown tasks.} In the absence of knowledge of the true values associated with detector data, physics and detector simulations are often critical for supervised learning tasks, as they provide a means for producing the ground-truth labels necessary to train ML models$\,$\cite{Carleo:2019ptp}. Unfortunately, it is common for deep learning models to learn unwanted features from simulation, leading to poor performance generalization of simulation-trained models to real data, otherwise known as Sim2Real gaps$\,$\cite{9398246}. Reference \cite{Schwartz:2021ftp} discusses ways to overcome this, one way being to train on real data. Our work takes this approach in the Migdal In Galactic Dark mAtter expLoration (MIGDAL) experiment's$\,$\cite{Araujo:2022wjh} search for the Migdal effect.

\begin{figure}[htbp]
\centering
\includegraphics[width=0.45\textwidth]{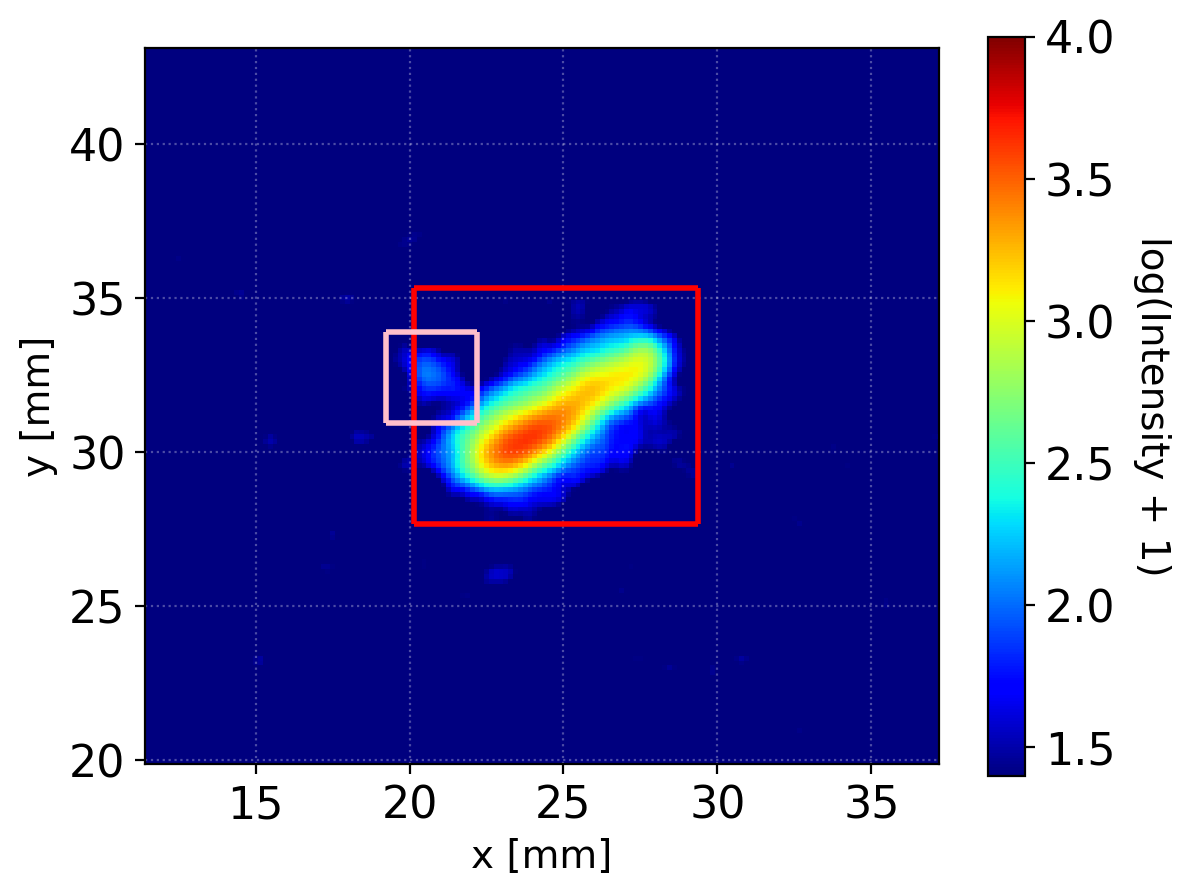}
\caption{Orca-Quest camera image postprocessed with $4\times 4$ pixel binning and Gaussian smoothing recorded in the presence of both an $^{55}$Fe calibration x-ray source and neutrons from the D-D generator, with YOLOv8's bounding box predictions shown. This event is illustrative of the characteristic 2D Migdal effect topology where a NR (red box; $\sim$\SI{310}{keV} visible energy) and ER (pink box; $\sim$\SI{6.0}{keV} visible energy) appear to share the same vertex. In actually, this event is not the Migdal effect, but rather the coincidental occurrence of an ER spatially overlapping with an NR within the \SI{8.3}{ms} exposure window of the camera. \label{fig:migtop}}
\end{figure}

MIGDAL is a neutron scattering rare event search experiment with the goal to detect and measure the Migdal effect for the first time$\,$\cite{migdal1939ionizatsiya,migdal1977qualitative}. In nuclear scattering, this effect arises when the sudden displacement of a recoiling nucleus induces the low-probability emission of an atomic electron, leading to a characteristic topology that consists of a nuclear recoil (NR) and electron recoil (ER) sharing a common vertex. Figure \ref{fig:migtop} shows an optical image of this topological signature recorded in the presence of an $^{55}$Fe x-ray calibration source inside the MIGDAL detector and neutrons from a deuterium-deuterium (D-D) generator. From this image, it is evident that the ER and NR portions of the event incur significant overlap, and that the NR portion of the track is orders of magnitude more intense than the ER. While it might seem paradoxical to train a deep learning model on real data for a rare event search, the Migdal effect can be treated as a composite signal consisting of an ER and NR. In isolation, ERs and NRs are commonly observed in the MIGDAL experiment, so if an object detection model can be trained on real data to reliably identify and localize ERs and NRs, then that information could be used for Migdal effect event selection. 

Our work, then, expands on the existing body of deep learning applications in nuclear and high energy physics in two ways. First, we train an object detection algorithm on \textit{real data} and apply it to a rare event search. Specifically, we train the YOLOv8 object detection algorithm on an abundance of real data spanning nine classes of events, with particular emphasis on ERs and NRs. We evaluate the trained model on a large sample of real data and use the spatial separation between identified pairs of ERs and NRs to optimize the selection of events consistent with the characteristic Migdal effect topology. Second, we demonstrate that our technique for identifying Migdal effect events is fast enough to perform our rare event search in real-time\footnote{Faster than the peak \SI{120}{fps} acquisition rate of $2048\times 1152$ pixel-frames in the CMOS camera readout of the MIGDAL experiment.} on a consumer-grade desktop PC, thereby enabling fully online Migdal effect searches with relatively modest hardware requirements.

The significant spatial overlap between the ER and NR portions of the Migdal effect topology make it inherently challenging to detect, thus our search is an ideal example to test the efficacy of our object detection approach that uses data-driven training to detect a composite rare event signal. We structure the rest of this work as follows:
In Sec.$\,$\ref{sec:Overview}, we elaborate on the Migdal effect and MIGDAL experiment, placing particular emphasis on the CMOS camera readout system, as YOLOv8 analyzes data from this system. Section \ref{sec:YOLOpipeline} introduces YOLOv8, our procedures for labeling data and training YOLOv8, and all steps of the raw data processing and analysis procedure that are automated by our pipeline. We also present benchmark studies that demonstrate our pipeline's capability of identifying Migdal effect candidates from raw data at real-time speeds. Section \ref{sec:Performance} presents several simulation studies that quantify YOLOv8's multiple track detection performance. These include studies of Migdal detection efficiencies versus energy and spatial overlap between ERs and NRs, the latter of which is relevant for any application that involves vertex reconstruction. This section concludes with quantifying YOLOv8's background rejection and signal retention when applying Migdal search criteria to simulation. Section \ref{sec:Reporting} highlights quantities computed and reported by our pipeline online and in real-time, including those relevant to our rare event search. The section concludes with conducting a Migdal effect search on a large sample of recorded CMOS camera images, demonstrating the scale of the data reduction we achieve when using YOLOv8 to search for Migdal effect candidates in real data. In Sec.$\,$\ref{sec:applications}, we discuss broader applications that could benefit from our approach of training object detection algorithms on real data to improve composite-signal reconstruction, emphasizing the analogy between our approach for Migdal effect reconstruction and decay vertex reconstruction. In particular, we discuss ways our approach could be applied to neutrinoless double beta decay experiments, and exotic nuclear decay reconstructions. Finally, we summarize our key findings in Sec.$\,$\ref{sec:summary}.

\section{Overview of the MIGDAL experiment}
\label{sec:Overview}

\begin{figure*}[htbp]
\centering
\includegraphics[width=0.9\textwidth]{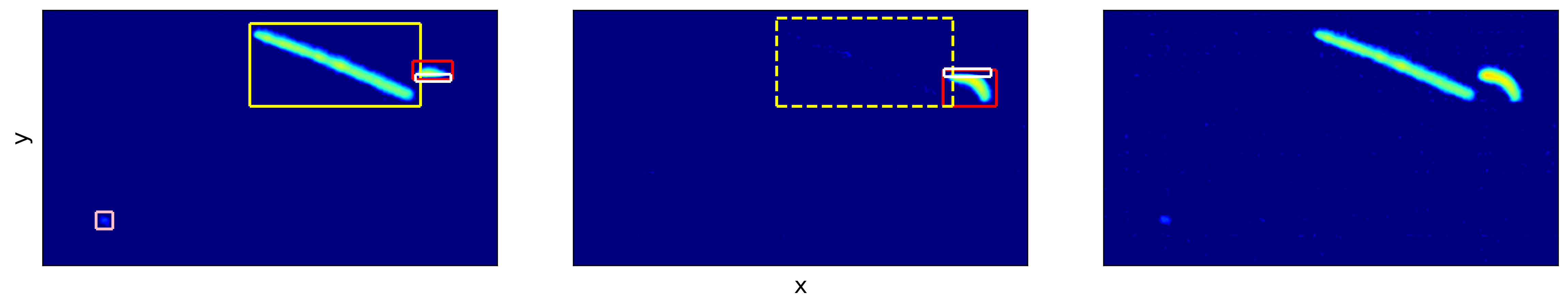}
\caption{Two consecutive camera frames at \SI{20}{ms} exposure (left and middle panels) and their sum (right) visualized with a logarithmic intensity scale. The pink, yellow, and red bounding boxes represent YOLOv8's bounding box predictions for an ER, a proton, and a NR, respectively (see Sec.$\,$\ref{sec:YOLOpipeline} for YOLO pipeline details). The white bounding boxes in the left and middle panels show YOLOv8's prediction that the rolling shutter clipped the NR and also estimate where the track was clipped. Summing these two frames together recovers the clipped NR at the expense of signal to noise. The faint long track inside the yellow dashed bounding box shows an example of a proton ghost.\label{fig:RS}}
\end{figure*}

Although the Migdal effect was predicted decades ago$\,$\cite{migdal1939ionizatsiya,migdal1977qualitative}, its relevance for the dark matter (DM) search community was firmly established only in 2017 -- with the derivation of an explicit relationship between the Migdal electron probability and ER and NR energies in Ref.$\,$\cite{Ibe:2017yqa}. While the Migdal effect is a very rare process, this derivation elevated the Migdal effect in nuclear scattering to an attractive process for enhancing the sensitivity of sub-GeV DM searches. In this DM mass regime, detected energies from Migdal electrons can far exceed those from DM-induced NRs, enabling the indirect detection of subthreshold NRs via the Migdal effect. Several experiments have since published DM scattering limits exploiting the Migdal effect to enhance their sensitivity to light DM$\,$\cite{LUX:2018akb,EDELWEISS:2019vjv,CDEX:2019hzn,XENON:2019zpr,SENSEI:2020dpa,COSINE-100:2021poy,EDELWEISS:2022ktt,DarkSide:2022dhx,XMASS:2022tkr,LZ:2023poo,PandaX:2023xgl,SuperCDMS:2023sql,SENSEI:2023zdf} but, as of this writing, there has been no experimental confirmation of the Migdal effect in nuclear scattering$\,$\cite{Xu:2023wev}.

\subsection{The MIGDAL experiment}
\label{subsec:detector}

The primary goal of the MIGDAL experiment is to make the first direct detection and measurement of the Migdal effect in nuclear scattering, which will be used to test theoretical predictions in Refs.$\,$\cite{Ibe:2017yqa,Cox:2022ekg}. To achieve sufficient statistics, the experiment uses a commercial deuterium-deuterium (D-D) fusion generator from Adelphi Technology Inc. The generator provides an approximately monoenergetic (\SI{2.5}{MeV}) source of neutrons with a nominal isotropic rate of $10^9\,\text{neutrons/s}$ incident on an optical time projection chamber (OTPC) filled with \SI{50}{Torr} CF$_4$ gas. Particle interactions with the CF$_4$ in the active volume produce a primary (S1) scintillation signal, and ionization amplification through a double glass-gas electron multiplier (GEM)$\,$\cite{Takahashi:2013rea} layer with a gain of $\mathcal{O}(10^5)$, produce a secondary (S2) scintillation signal. Both signals are recorded by a 3-inch Hamamatsu R11410 photomultiplier tube (PMT) readout that is capable of determining the absolute z position of recoils from the time difference between the S1 and S2 signals. The latter signal is also imaged by a Hamamatsu ORCA-Quest qCMOS camera (OQC)$\,$\cite{orca}. The generator and OTPC are located at NILE/ISIS at the STFC Rutherford Appleton Laboratory in the UK.  To date, the MIGDAL experiment has collected D-D-generator data over two dedicated science runs spanning several weeks. We briefly introduce the MIGDAL OTPC and the OQC readout, since our pipeline is designed for this readout.

\textbf{MIGDAL detector} At the core of the experiment is a \SI{110}{cm^3} OTPC with combined optical and electronic readouts, allowing for full 3D reconstructions of particle tracks$\,$\cite{ITO,Tilly:2023fhw}. A collimator and shield separate the D-D generator from the OTPC, with the neutrons incident in the +$x$ direction of the OTPC. To minimize diffusion while maintaining reasonable neutron interaction rates, the drift region of the detector spans \SI{3.0}{cm}, with a \SI{200}{V/cm} electric field applied between a cathode mesh and the first GEM; \SI{600}{V/cm} in the transfer gap between the two GEMs; and \SI{400}{V/cm} in the induction gap between the second GEM and the Indium Tin Oxide (ITO) anode plane. The ITO plane is segmented into 120 charge readout strips of \SI{833}{\um} pitch that together provide $x$-$z$ projections of events$\,$\cite{ITO}. Pairs of strips 60 strips apart are connected to 60 charge amplifiers and 8-bit digitizer channels, operating at a \SI{500}{MHz} sampling rate and providing a $z$ granularity of \SI{260}{\um/sample}. The OQC images an $8.0\times \SI{4.5}{cm^2}$ region in the transverse $x$-$y$ plane, leading to an observable active volume of $\SI{8.0}{cm}\times \SI{4.5}{cm}\times \SI{3.0}{cm}$. An external \SI{80}{MBq} $^{55}$Fe x-ray source is attached to the OTPC on a movable shaft and can be remotely deployed for energy calibrations. This electron capture source emits \SI{5.9}{keV} K-$\alpha$ and \SI{6.5}{keV} K-$\beta$ x-rays from $^{55}$Mn fluorescence, which irradiate the active volume with relative intensities of around $87\%$ and $13\%$. More detailed descriptions of the experiment and detector operations can be found in Refs. \cite{Araujo:2022wjh,Knights:2024wjh}.

\textbf{Camera readout}
The OQC images the S2 light from the output of the second GEM through an EHD‐25085-C F0.85 lens$\,$\cite{ehdlens}. As such, throughout this work, we reconstruct track energies using this light signal and report ``electron-equivalent" energies, which we denote with an ee-subscript (e.g.$\,\rm keV_{ee}$). The OQC sensor measures $4096\times 2304$ pixels, corresponding to a \SI{20}{\um} pixel scale projected onto our field of view on the second GEM. The sensor is Peltier-cooled to $-20^{\circ}$C to limit dark current to negligible levels relative to readout noise. The camera operates with a continuous rolling-shutter with two acquisition modes designated as ``Standard" and ``Ultra quiet". With Hamamatsu’s proprietary CoaXPress cable to interface with the readout PC, images are recorded at up to \SI{120}{fps} in Standard mode at the expense of signal to noise (0.43e RMS), while the Ultra quiet mode provides ultralow noise (0.27e RMS) at the expense of readout speed (\SI{5}{fps}). Both acquisition modes are used in the experiment, with Standard being used for recording neutrons from the D-D generator, and Ultra quiet primarily used for energy calibrations with the $^{55}$Fe source.

A consequence of the continuous rolling shutter mode is the possibility of tracks being clipped, with a portion of the track in one frame and the remainder the track in the next. In these cases, tracks can be recovered by stitching the two camera frames together as shown in Fig.$\,$\ref{fig:RS}. Image lag or ``ghosting" -- where a dimmed version of an event in a given frame persists into subsequent frames -- is another effect that is well known to occur in CMOS cameras$\,$\cite{imagelag} and important to identify in our analyses (see Sec.$\,$\ref{subsec:pipeline}). Moving forward, we will call image lagged events ``ghosts".

To reduce data volume while retaining good spatial resolution, we recorded data using an on-chip $2\times 2$ binning in both science runs, reducing the image size to $2048\times 1152$ 16-bit pixels.\footnote{A 2x2 binned pixel images a \SI{40}{\um} region projected on the GEMs, which have a hole pitch of \SI{270}{\um}.} With this reduction, the raw image data collection rate at \SI{120}{fps} acquisition is about \SI{2}{TB/hour}. Logistical issues during Science Run 1 required the use of USB 3.1 to interface the camera with the readout PC, limiting acquisition rates to \SI{50}{fps} over the entire run. In Science Run 2 we recorded neutron data at the full \SI{120}{fps} using the CoaXPress cable.

\subsection{Considerations for Migdal effect searches}
\label{subsec:constraints}

A key goal of this work is to demonstrate YOLOv8's performance in selecting Migdal-like event topologies in the 2D OQC images, which is a critical step in our search. We emphasize, however, that the OQC alone is not sufficient for confirming the Migdal effect and that information from the ITO and PMT subsystems is needed to provide the full event reconstruction necessary to make a confirmation. Besides providing the 3rd dimension of the event, the \SI{2}{ns} timing of the ITO is crucial for ruling out coincident ER-NR pairs unrelated to the Migdal effect (e.g. Fig.$\,$\ref{fig:migtop}), which can occur during the \SI{8.33}{ms} camera exposure. Therefore, selection criteria based on information integrated from all detector subsystems will be required to define the final set of Migdal candidates.

Reference \cite{Araujo:2022wjh} sets conservative region of interest (ROI) thresholds for the MIGDAL experiment's Migdal effect search of $\SI{15}{keV_{ee}}\geq E_\mathrm{ER}\geq \SI{5}{keV_{ee}}$, $L_\mathrm{ER}\geq\SI{4}{mm}$, and $E_\mathrm{NR}\geq\SI{60}{keV_{ee}}$.\footnote{This is an approximation of the nuclear recoil threshold after ionization quenching, as Ref.$\,$\cite{Araujo:2022wjh} quotes the threshold in terms of recoil energy: $E_\mathrm{NR}\geq\SI{100}{keV_{r}}$.} Here $L_\mathrm{ER}$ is the 2D length of the ER track. These thresholds are dictated by ensuring (1) that ERs are sufficiently long to resolve the ER head outside of the NR penumbra, and (2) that the NR energy is high enough that there is no ambiguity in its particle identification. As is discussed in Appendix$\,$\ref{subsec:single_trackSim}, YOLOv8 does an excellent job detecting NRs down to \SI{20}{keV_{ee}} and ERs down to about \SI{3.6}{keV_{ee}}, so we are sensitive to detections well below this conservative threshold.

\section{The end-to-end pipeline}
\label{sec:YOLOpipeline}
We are interested in using YOLOv8 as a tool to identify topologies consistent with the Migdal effect in 2D OQC data. One approach to achieve this goal would be to train YOLOv8 to directly identify Migdal effect topologies on simulated images. This approach comes with the notable downside of the often-observed poor-performance generalization when applying ML models trained on simulation to real data. To circumvent this Sim2Real gap, we instead employ a fully data-driven approach to searching for the Migdal effect in the OQC where we reframe the Migdal search as a search for pairs of ERs and NRs within close proximity of one-another, including those that spatially overlap. Framing the Migdal search in this way allows us to train YOLOv8 on an abundance of measured ER and NR tracks observed over the course of the two science runs.

Here we detail all steps of the automated image processing and rare event search analysis pipeline. We begin with a brief overview of YOLOv8 and then describe our procedures for labeling data and training YOLOv8. After this, we detail the steps of our pipeline to process raw image data and use the extracted information to search for Migdal effect candidates. We conclude this section with a benchmark study demonstrating the end-to-end processing and analysis speeds of our pipeline on our readout PC.

\subsection{YOLOv8}
\label{subsec:YOLOv8}
YOLOv8 is a Pytorch-based$\,$\cite{paszke2019pytorch} open-source model released by Ultralytics in January 2023. Here we briefly describe important details of YOLOv8's loss function and the various YOLO models present in the Ultralytics YOLOv8 package. Specific details on YOLOv8's architecture can be found in Ref.$\,$\cite{terven2023comprehensive}. 

A useful metric for quantifying the spatial overlap between bounding boxes is known as intersection-over-union (IoU). Given two bounding boxes $B_1$ and $B_2$, their IoU is computed as

\begin{align}
\label{eq:IOU}
    \mathrm{IoU}(B_1,B_2)&\equiv\frac{|B_1\cap B_2|}{|B_1\cup B_2|},
\end{align}
where $|B_1\cap B_2|$ and $|B_1\cup B_2|$ denote the area of overlap of the regions of intersection and union of $B_1$ and $B_2$, respectively.
Intuitively, IoU seems like a good quantity to optimize when training an object detection algorithm, as it is scale invariant; however, by definition, it is always zero when there is no overlap between a predicted bounding box and the ground truth bounding box. The introduction of generalized intersection-over-union (GIoU)$\,$\cite{rezatofighi2019generalized} solved this problem by adding an additional penalty term to Eq.$\,$(\ref{eq:IOU}) that is related to the distance between the two bounding boxes, making GIoU a viable metric for a bounding box loss function that can be optimized when training. Complete intersection-over-union IoU (CIoU) was later introduced and takes into account the aspect ratio of the bounding boxes, which led to a substantial improvement in average precision scores on the PASCAL VOC 2007 dataset compared to GIoU$\,$\cite{zheng2020distance}.

All together, the YOLOv8 loss function consists of three terms: (1) CIoU loss; (2) distribution focal loss (DFL)$\,$\cite{li2020generalized}; and (3) binary cross entropy loss with a sigmoid function applied to the class prediction associated with a bounding box. DFL also aims to optimize bounding box regression but, unlike CIoU, DFL predicts the distribution of possible bounding box offsets, thereby reducing the uncertainty in bounding box location. Terms (1) and (2) together, then, optimize YOLOv8's localization performance, while term (3) optimizes bounding box classification. For a given image, YOLOv8's loss function achieves a minimum when all predicted bounding boxes and their associated classifications agree exactly with the ground truth labels for that image. The exact functional form for YOLOv8's loss function can be found in Ref.$\,$\cite{2023arXiv230509972R}.

Out of the box, the Ultralytics YOLOv8 package contains five model architectures labeled in order of smallest to largest as ``n", ``s", ``m", ``l", and ``x". When evaluated on the MS COCO 2017 test set$\,$\cite{lin2014microsoft}, model ``x" was found to outperform all previous releases of YOLO$\,$\cite{yolov8}. In addition to object detection, the YOLOv8 package also contains models for image classification (that can be easily adapted to regression), key point detection, and instance segmentation, so the YOLOv8 package is capable of performing all of the tasks highlighted in Fig.$\,$\ref{fig:CV}. Hereafter, we drop the ``v8" designation and refer to YOLOv8 simply as YOLO.

\begin{figure*}[htbp]
\centering
\includegraphics[width=0.8\textwidth]{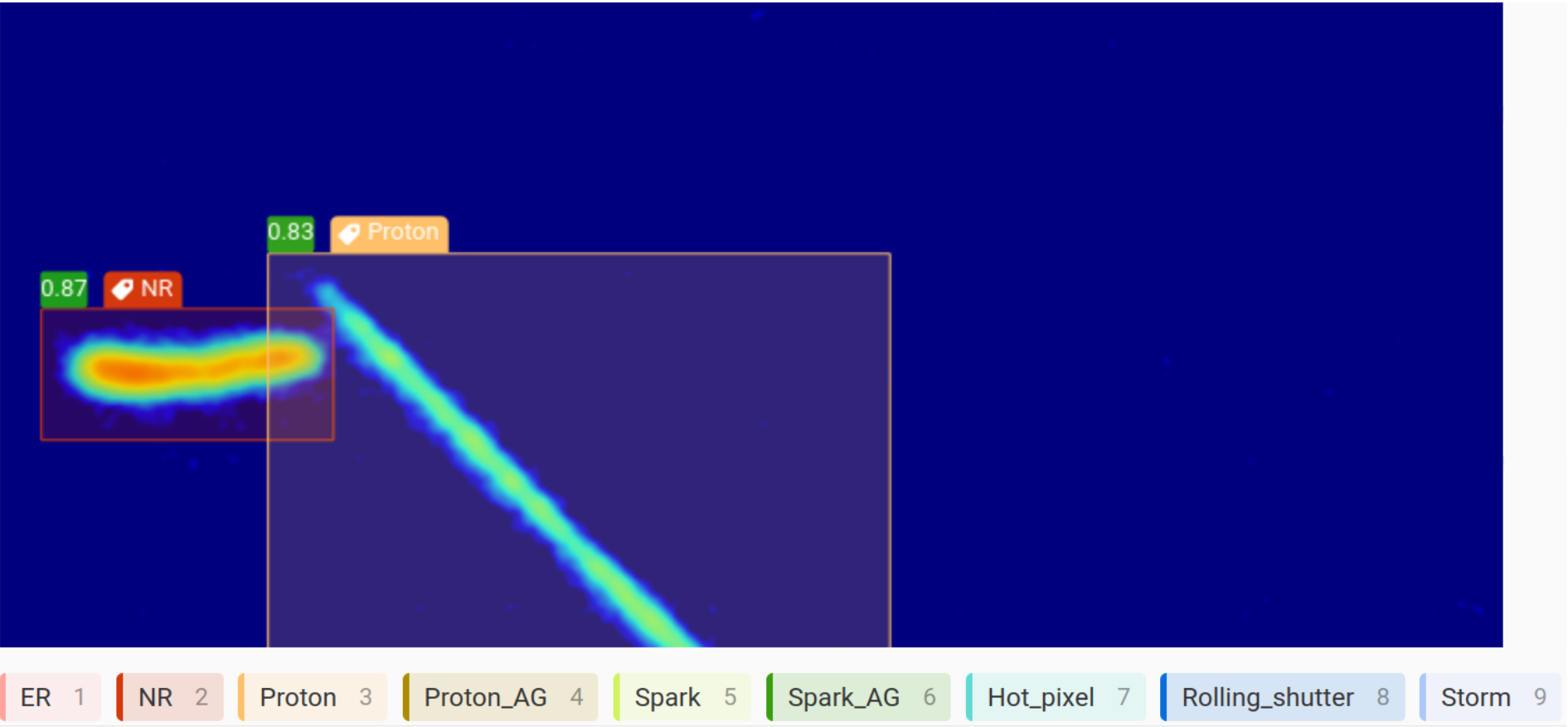}
\caption{Snippet of the Label Studio front-end interface showing a logarithmic-intensity-scale image with preannotated bounding boxes and associated classification labels and scores, produced by a YOLO model pretrained on the training data from Table$\,$\ref{tab:i}.\label{fig:preannotation}}
\end{figure*}

\subsection{Data labeling and training YOLO}
Of the five previously mentioned YOLO model architectures, we train model ``m" from scratch using measured OQC data, as we find this model to strike the best balance between object detection performance and inference speed for our real-time application. Naturally, training on measurement with human-labeling comes with the trade-off of model performance being limited to human-level performance. Reference \cite{9455314} shows that human-level performance limitations can potentially be overcome by augmenting real data training samples with high quality labeled simulation. We test this later (Secs.$\,$\ref{subsec:SimResults}-\ref{subsec:coincrej}) and find that augmenting our real data training set with simulation improves performance in several metrics. Human labeling also comes with the downside of the labeling process being labor intensive, however once a model is sufficiently trained, we can use automated preannotations to significantly speed up the data labeling process. Biases in drawing bounding boxes, as well as assigning class labels to events with ambiguous particle identification in the 2D OQC images, such as short-length sub-\SI{10}{keV_{ee}} ER and NR tracks, can also be problematic. In our case, all training data were labeled by a single user, which may reduce variation in the geometries of labeled bounding boxes, but may also amplify biases in assigning class labels to ambiguous tracks. Using physically motivated metrics to assign track labels can help reduce bias in these scenarios.

We label data using an open-source data labeling platform called Label Studio$\,$\cite{LabelStudio}, that provides data labeling templates for many ML applications, including several computer vision tasks. Before uploading to Label Studio, each training image undergoes preprocessing steps that include dark subtraction, 4$\times$4 binning and Gaussian smoothing, and finally, conversion to PNG format using an empirically determined fixed logarithmic intensity dynamic range.\footnote{See steps 1-3 of our analysis pipeline described in Sec. \ref{subsec:pipeline} for further details about the preprocessing procedure.} The PNG images are then uploaded to Label Studio where we use an object detection template that provides a click-and-drag interface for bounding box labeling. After annotating an image with all appropriate bounding boxes, the template generates a label text file for the image in the normalized $xywh$ format -- a commonly used format for bounding box labels. Specifically, each ground truth bounding box, $B_\mathrm{t}$, in the labeled image corresponds to a single line in the label text file that contains the following contents in this order: (i) the class label $y_{\mathrm{t}}$; (ii) and (iii) the coordinates of the $x$ and $y$ centroids of $B_\mathrm{t}$: $\overline{b}_{x,\mathrm{t}}$ and $\overline{b}_{y,\mathrm{t}}$, respectively; and (iv) and (v) the width and height of $B_\mathrm{t}$: $w_\mathrm{t}$ and $h_\mathrm{t}$, respectively. We denote bounding boxes predicted by YOLO as $B_\mathrm{p}$.

\begin{table}[b!]
\caption{\label{tab:i}%
Class name and total number of instances (number of bounding boxes) of each of the nine classes in our labeled training data. All labeled data were extracted from measurements recorded during the two science runs. Frames containing multiple instances of tracks (including those of the same class) are included in this sample.
}
\begin{ruledtabular}
\begin{tabular}{lr}
Class Label & $N_\mathrm{instances}$\\
\hline
Electron recoil or NR ghost\footnote{Due to the similar d$E$/d$x$ signatures of NR ghost tracks and ER tracks, we label NR ghosts as ERs and train YOLO to initially classify NR ghosts as ERs. NR ghosts are flagged at a later stage in our pipeline (see Sec.$\,$\ref{subsec:pipeline}).} & 4,396\\
Hot pixel & 130\\
Nuclear recoil & 3,840\\
Proton or alpha & 300\\
Proton ghost or alpha ghost & 63\\
Rolling shutter clip & 539\\
Spark & 135\\
Spark ghost & 172\\
High occupancy shower\footnote{These events consist of dense pileups of ER tracks that cover significant portions of an image frame. We are unable to resolve individual tracks among the pileup, so we train YOLO to identify the presence of these showers so we can reject them from our analysis. These showers are relatively uncommon, typically occurring in fewer than one in one thousand frames at \SI{120}{fps} acquisition.} & 58\\\hline
Total & 9,633\\
\end{tabular}
\end{ruledtabular}
\end{table}

Label Studio has functionality for automated preannotations through ML-assisted labeling. To utilize this feature, we first train YOLO using the preprocessed PNG images. After YOLO has been sufficiently trained, we inject the trained model weights to a custom-written ML back-end plugin that interacts with Label Studio's object detection template. This plugin evaluates our trained YOLO model to preannotate uploaded PNG images with editable bounding boxes in the Label Studio front-end. Figure \ref{fig:preannotation} shows an example of the automated preannotations placed on a NR candidate and a proton candidate in the Label Studio interface. Generally speaking, as model performance improves with more training, fewer manual adjustments are required on preannotated data, thereby streamlining the labeling process.

Table$\,$\ref{tab:i} shows the class breakdown of human-labeled data used to train YOLO. Since Migdal searches are our application of interest, the overwhelming majority of tracks we train on are ERs and NRs, as these are the constituents of Migdal effect topologies. We split the set of training frames with their associated labels into a $70\%/30\%$ training/validation split. In each training epoch, we optimize YOLO's model weights using stochastic gradient descent with momentum and weight decay$\,$\cite{SGD} and employ early stopping$\,$\cite{estop} where we terminate training after 25 successive epochs with no improvement in the mAP@50:95:5 metric (mean average precision over IoU thresholds of 0.5 to 0.95 in steps of 0.05)$\,$\cite{metrics}. When training YOLO using a training set that is augmented with additional simulation (as in Sec.$\,$\ref{subsec:SimResults}), we follow this same training procedure.

\subsection{YOLO-based data processing and analysis pipeline}
\label{subsec:pipeline}
Next, we describe the automated pipeline for processing and analyzing images batch-by-batch. This pipeline is integrated with our experiment’s data acquisition software and begins processing batches of 200 camera frames as they are written to disk. The steps of our pipeline are as follows: 

\begin{enumerate}
\item \textbf{Dark subtraction}: We pregenerate two $2048\times 1152$ master dark frames: one that is used for Ultra quiet data acquisition and the other for Standard acquisition. These frames are generated from two dedicated dark runs where, in each run, 800 camera frames are recorded with the rest of the detector powered off. Master dark frames are created from the mean intensity of each pixel over the 800 frames, with sigma clipping at a $5\sigma$ tolerance applied.\footnote{Sigma clipping is used to remove anomalous pixel intensities when creating master dark frames. First, we compute the mean and standard deviation intensity of each image-pixel over the 800 dark frames. Then, we mask all pixels with intensities greater than $5\sigma$ of the mean and repeat the process until there are no remaining pixels to mask.} During a run, the appropriate master dark frame is subtracted from each batch of recorded images.
\item \textbf{Downsample and Gaussian smoothing}: Each batch of images is downsampled using $4\times 4$ binning, yielding 200 frames with $512\times 288$ pixels. A $9\times 9$ Gaussian smoothing kernel with $\sigma_x=\sigma_y=4/3$ is then applied to the stack of frames. For performance, we use Pytorch to perform both the downsampling and Gaussian smoothing on a GPU.
\item \textbf{Convert images to PNG format}\footnote{This step is only necessary because we trained YOLO on PNG images. In principle, YOLO could be trained on processed $512\times 288$ intensity arrays which would render this step irrelevant.}: Each camera frame is converted into a PNG file to be passed as input into YOLO. To best capture the large dynamic range required for simultaneously elucidating ERs and NRs, we use an empirically determined fixed logarithmic intensity scale with a minimum of $\log_{10}(I+1)=1.4$ and maximum of $\log_{10}(I+1)=4$, where $I$ is the intensity of a $4\times 4$-binned, Gaussian-smoothed pixel.
\item \textbf{Pass the PNG images into YOLO and store the coordinates and classifications of each bounding box}: The trained version of YOLO simultaneously localizes and classifies (using the nine class labels from Table$\,$\ref{tab:i}) all objects in each PNG image. The bounding box pixel coordinates on the PNG images are mapped back to the original 200 downsampled and Gaussian smoothed $512\times 288$ frames. YOLO runs on a GPU for performance.
\item \textbf{Extract physical information from bounding box contents}: A key benefit of YOLO classifying and localizing events at the same time is that these classifications can immediately be used to decide which bounding boxes to extract physical content from. We do not want to waste computational resources performing fits on a spark event, for example. Given this, for each bounding box identified as an ER, NR, or proton/alpha, we compute the following quantities: (i) track intensity/energy, (ii) track length, (iii) axial angle, (iv) for ERs and NRs only: head/tail asymmetries and vector angles (Sec.$\,$\ref{subsec:online}). Computations of track energies require intensity corrections due to vignetting (Appendix$\,$\ref{sec:A1}), which is done using data from $^{55}\rm Fe$ calibration runs.

\item \textbf{Flag and reject NR ghosts}: NR ghosts have the potential to be a significant background with d$E$/d$x$ signatures that mimic ERs, so they must be flagged and rejected to reduce false positives in a Migdal search. Since YOLO is trained to initially identify NR ghosts as ERs, we flag NR ghosts by first computing the IoU overlap between each ER in a given frame and each NR in the previous frame. Then, if an ER in the given frame and NR in the previous frame have IoU overlap greater than 0 and the peak pixel intensity of the $4\times 4$-binned Gaussian smoothed NR is greater than an empirically determined threshold of \SI{150}{ADU}, we flag that initially identified-ER as a NR ghost.

\item\textbf{Identify and analyze all unique fiducialized ER-NR pairs}: After rejecting NR ghosts, we identify all frames containing fiducialized pairs of ERs and NRs. For each unique pair, we compute the IoU overlap between the predicted ER and NR bounding boxes,  IoU($B_\mathrm{p}^\mathrm{ER}$,$B_\mathrm{p}^\mathrm{NR}$), and the distance between the centroid of $B_\mathrm{p}^\mathrm{ER}$ and point of highest intensity of $B_\mathrm{p}^\mathrm{NR}$. This is a critical step in identifying Migdal effect candidates that will be discussed in more detail in Secs.$\,$\ref{sec:Performance} and \ref{subsec:migskim}.

\item \textbf{Save extracted information}: For each image batch, we write an output file with its contents indexed by bounding-box, thereby converting frame-indexed data to event-indexed data. In addition to the physical information extracted from each bounding box, we also store the coordinates of the corners of the bounding box, classifications with associated confidence scores, and metadata such as image timestamps, run IDs, file and frame indices, and, optionally, the coordinates and intensities of each pixel within each bounding box. Pairs of ERs and NRs showing up on the same frame that satisfy user-defined Migdal search criteria are flagged as candidate events and are saved separately alongside of the event-indexed data. Candidate event files are indexed by ER-NR pair and include pair-specific quantities such as the IoU overlap between the ER and NR in the pair, as well as the distance between the centroid of the ER bounding box and point of highest intensity within the NR bounding box.
 
\end{enumerate}

During a run, after the pipeline performs the above steps, each event-indexed processed file is then added to a temporary database that reports event rates, track energy versus length distributions of several event species, and NR energy spectra in a live display that updates in real-time (see Fig.$\,$\ref{fig:realtime} in Sec.$\,$\ref{sec:Reporting}). Counts of candidate ER-NR pairs are also enumerated in the live display. At the end of the run, all processed files are transferred to an offline computing server.

\subsection{Benchmarking end-to-end processing speed}
To verify our claim that this pipeline is capable of processing and performing a rare event search on OQC data online and in real-time, with modest hardware requirements, we developed a benchmark script available at$\,$\cite{schueler_2024_12628437}. This script can be run on any PC but its configuration parameters have been optimized for the MIGDAL readout PC which uses an Intel Core i9-10900X CPU and a single slot NVIDIA RTX A4000 graphics card. On the MIGDAL readout PC, the script runs two processes:\newline\newline
\textbf{Process 1}: This process is run as three parallel subprocesses that each read in a batch of 200 raw images, perform steps 1-3 of the pipeline, and save the processed image batch and 200 PNG images. Each iteration of Process 1 therefore processes 600 raw images.\newline\newline
\textbf{Process 2}: This process performs steps 4-8 of the pipeline using the outputs of Process 1 as input. When this process completes, it deletes the processed image batch and PNG images generated in Process 1.\newline

\begin{figure}[htbp]
\centering
\includegraphics[width=0.48\textwidth]{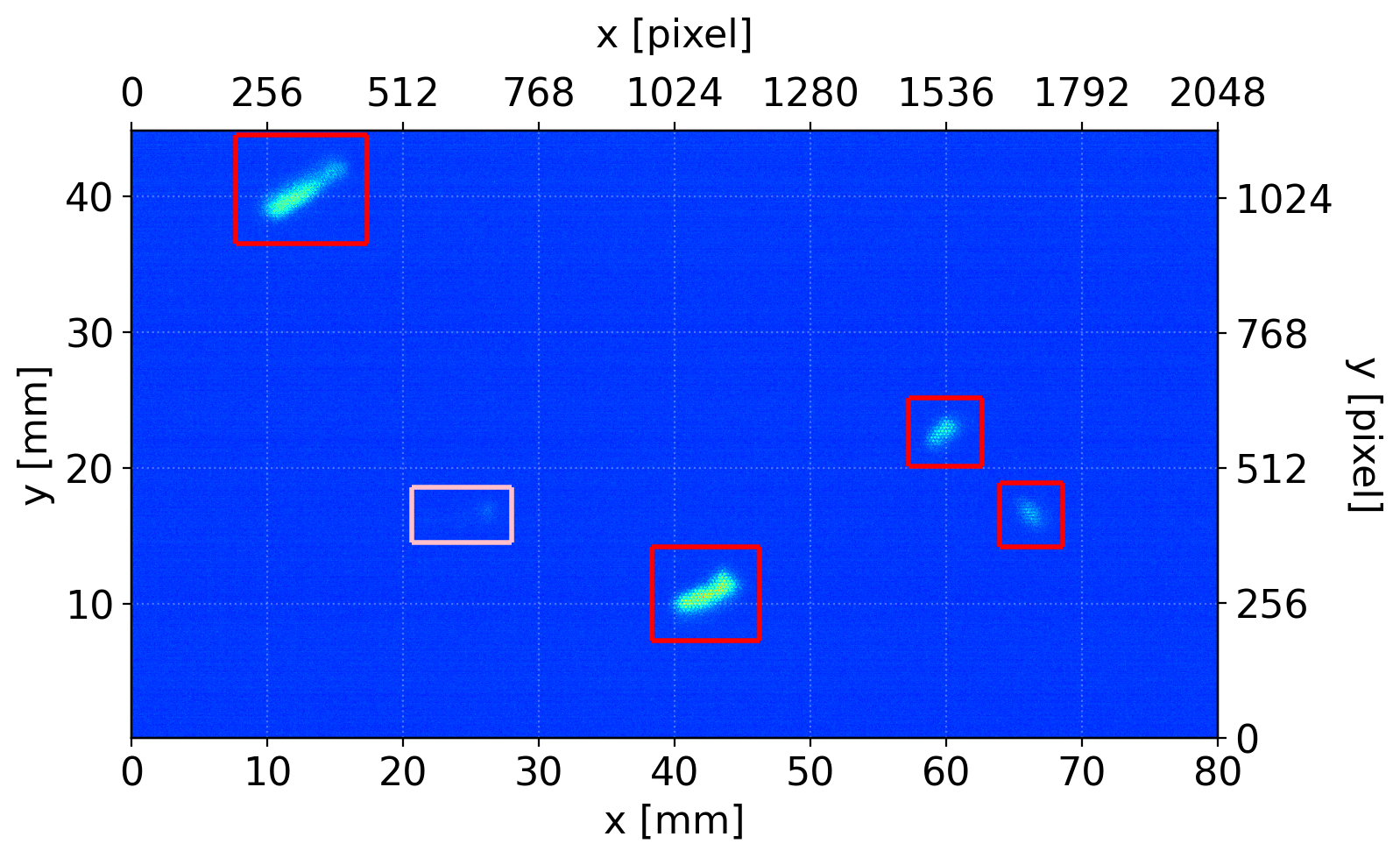}
\caption{Unprocessed frame (linear intensity scale) used for the high occupancy batch with YOLO's bounding box predictions shown. The pink bounding box is an ER prediction and the red bounding boxes are NR predictions.\label{fig:highocc}}
\end{figure}

To compute the end-to-end pipeline processing time, timestamps are saved for each image batch at the beginning of Process 1 and at the end of Process 2. Each raw batch of images has a file size of around \SI{900}{MB}, so to avoid data storage burdens when benchmarking the processing time of the pipeline, we run the benchmark script repeatedly on a single batch of 200 images. We perform two benchmarks: one using a randomly selected batch of 200 images and the other using a custom high occupancy batch formed with 200 copies of the frame shown in Fig.$\,$\ref{fig:highocc}.

\begin{figure}[htbp]
\centering
\includegraphics[width=0.48\textwidth]{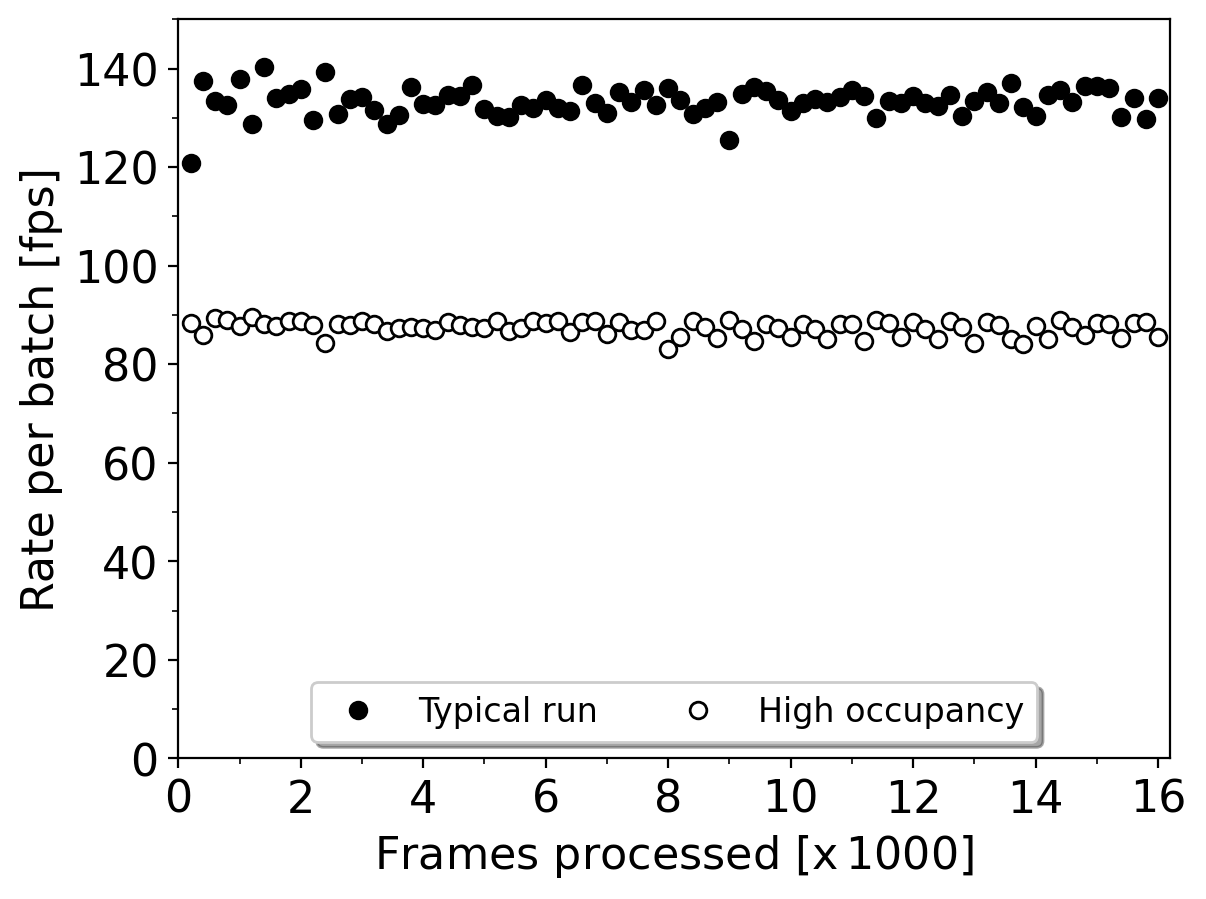}
\caption{Pipeline processing rate as a function of number of frames processed for a batch with events corresponding to a typical \SI{20}{ms} exposure run (filled points), and our custom high occupancy batch (unfilled points). Downsampling, which we parallelize into three subprocesses, is the performance bottleneck for the typical run sample, leading to a larger variance in when the inputs of Process 2 are generated, hence the larger spread in overall processing rates for this sample. \label{fig:benchmark}}
\end{figure}

Figure \ref{fig:benchmark} shows the results of this benchmark study evaluated over 16,000 images (80 batches). The filled points show the processing speed of a randomly selected batch from a typical \SI{20}{ms} exposure (\SI{50}{fps}) run in the presence of neutrons from the D-D generator. The event occupancy within frames in such a run is, on average, 2.4 times higher than an identical run with \SI{120}{fps} acquisition. Nevertheless, our benchmark shows that we consistently process and analyze these frames faster than \SI{120}{fps}. While our benchmark does not achieve \SI{120}{fps} on the high occupancy batch, we achieve an average processing rate near \SI{90}{fps}, indicating that our pipeline is able to process and analyze tracks at rates close to \SI{450}{Hz} in a higher occupancy environment. These benchmark studies not only show that we are able to process and analyze typical data online and in real-time, but also that our pipeline is capable of processing significantly higher track rates than the designed peak NR rate of MIGDAL$\,$\cite{Araujo:2022wjh}. 

\section{Pipeline performance}
\label{sec:Performance}
With the end-to-end speed of our pipeline established, we next evaluate the object detection performance of our implementation of YOLO. While we train YOLO on measurement, we need simulation to quantitatively evaluate YOLO's detection and localization performance. Appendix$\,$\ref{sec:AppendixSingleTrack} describes the key steps of producing realistic optical simulation, and validates YOLO's identification performance on simulated single-track frames consisting of either ERs or NRs. Since the Migdal effect can be treated as a composite signal consisting of an ER and NR, we focus here on analyses of YOLO's performance in identifying two tracks in a single frame, with an emphasis on Migdal candidate detection and background rejection. We perform these two-track analyses primarily on  ``hybrid" simulations, which consist of measured NR frames that are stitched together with simulated ERs to form either simulated Migdal events or simulated coincidence events.

An analysis solely using OQC data is limited by the relatively long \SI{8.33}{ms} exposure time, giving rise to ``accidental" coincidences where independent ERs and NRs can be recorded in the same camera frame. These accidental coincidences are the primary background in an OQC-only analysis. The much faster \SI{2}{ns} sampling of the ITO can and is used to reject these coincidences, but we can also use the fact that Migdal-like topologies consist of an ER and NR in close proximity to one another to reject the majority of accidental coincidences in OQC frames. Coincidences can also arise from background physical processes that generate correlated ER-NR pairs, however the background studies performed in Ref.$\,$\cite{Araujo:2022wjh} suggest that correlated coincidences satisfying our search ROI are rarer than the Migdal effect within our ROI, so we ignore these in our OQC studies. In this section, we quantify both YOLO's Migdal identification and accidental coincidence background rejection performance on simulation. Moving forward we simply refer to accidental coincidences as ``coincidences".

\begin{figure*}[htbp]
\centering
\includegraphics[width=0.85\textwidth]{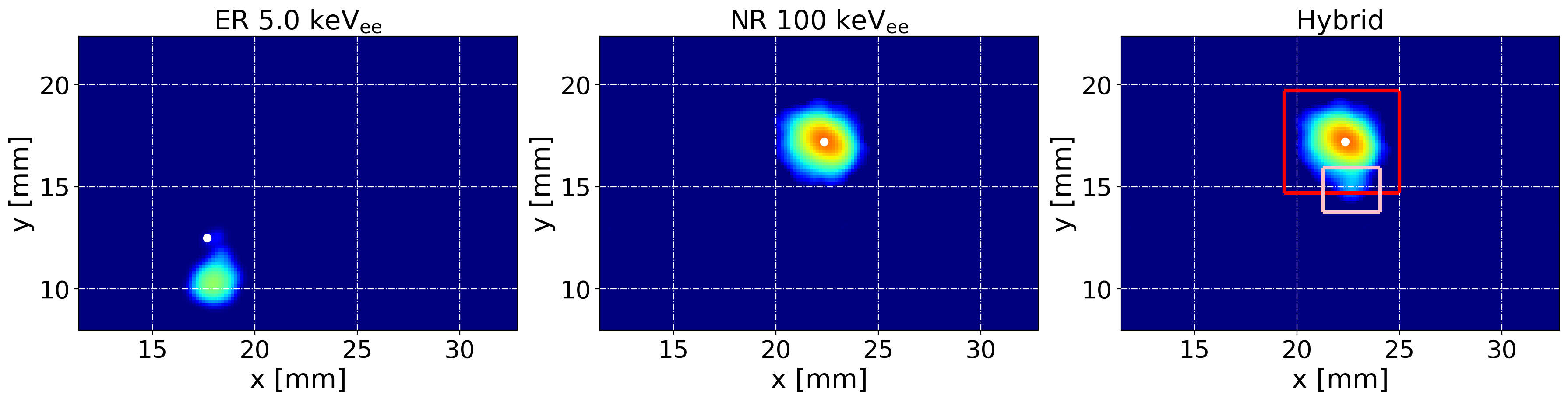}
\includegraphics[width=0.85\textwidth]{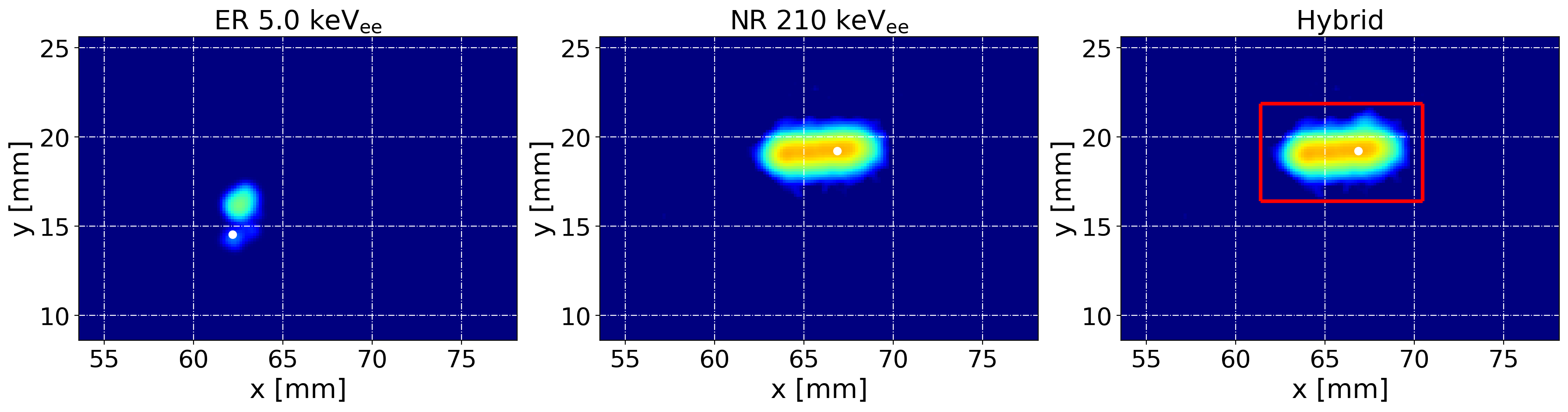}
\includegraphics[width=0.85\textwidth]{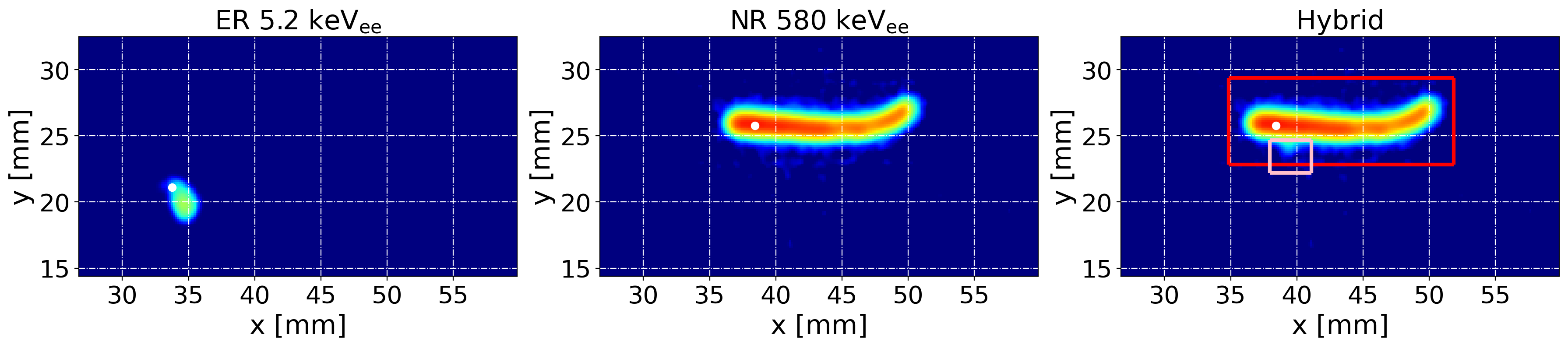}
\caption{Simulated ERs, measured NRs, and hybrids plotted on logarithmic intensity scales. Each row shows an example of stitching together the vertex of a simulated ER with the point of highest intensity of a measured NR to generate a hybrid Migdal event. Left column: Simulated ER tracks with gain scaling but no noise; truth ER vertices are shown in white. Middle column: Measured NRs with their point of highest intensity shown in white. Right column: hybrid Migdal formed by translating the ER so its vertex aligns with the point of highest intensity of the NR; vignetting scaling is applied to the ER once it's aligned with the NR. The bounding box predictions of YOLO trained on the Augment sample (Sec.$\,$\ref{subsec:SimResults}) are shown for the three hybrids with red (pink) boxes denoting NR (ER) predictions. The top and bottom rows show examples of positive detections. YOLO did not identify the ER in the middle-row hybrid indicating a negative detection. The fractions of ER significant pixels, $\rm f_{sigpix}$ (see Sec.$\,$\ref{subsec:SimResults2}), in the top, middle, and bottom-row hybrids are 0.42, 0.27, and 0.10, respectively. \label{fig:hybrid}}
\end{figure*}

\subsection{Description of samples}
\label{subsec:twotracksim}
We study Migdal and coincidence identification using both a purely simulated sample and a ``hybrid" simulation, where we start with \textit{measured} frames (real data) containing a single NR and add a single simulated ER to those frames -- recalling that the Migdal probability is so small that it is unlikely we will bias this study due to the NR sample containing real Migdal events. Specifically, we create the following four samples, where each frame consists of exactly one ground truth simulated ER and one NR:

\begin{enumerate}
\item \textbf{Pure simulation Migdal events}: Using a sample of 10,500 pure simulated ERs following a discrete uniform energy distribution varying in steps of \SI{0.2}{keV} between \SI{2.0}{keV} and \SI{6.0}{keV} inclusive, and the sample of simulated NRs described in Appendix$\,$\ref{subsec:single_trackSim}, we select around 27,000 NRs with $E_\text{NR}>\SI{60}{keV_{ee}}$. Then, for each NR frame, we shift the NR track to a random position on the readout, ensuring that no NR pixel falls within 20 pixels of any edge of the frame. After this, we select a random simulated ER track and translate the ER track so its truth vertex aligns with the pixel of highest intensity of the NR.\footnote{We approximate the NR vertex as the pixel of highest intensity.} Once both tracks are aligned, we follow the procedure described in Appendix$\,$\ref{subsec:ERsim} to perform gain scaling, vignetting scaling, and noise addition using a randomly selected dark-subtracted dark frame from a sample of 800 such frames. 

\item \textbf{Pure simulation coincidence events}: Selecting from the same set of pure simulated NRs and ERs, we repeat the procedure above except instead of aligning the ER truth vertex with the point of highest intensity of the NR, we separate the ER and NR vertices in both $x$ and $y$ drawing from separate random uniform distributions for each coordinate.

\item \textbf{Hybrid Migdal events}: We form these starting with around 27,000 measured frames, where YOLO identified a single NR with a reconstructed energy above \SI{60}{keV_{ee}} and nothing else. For each of these frames, a random ER from the sample of 10,500 simulated ERs previously described is selected and translated so its ground truth vertex matches the point of highest intensity of the measured NR. Gain scaling and vignetting scaling are then applied to the ER. Since the NR frame is from measurement, it already contains noise, so we do not add any noise to the simulated ER when stitching it to the NR frame. Furthermore, we do not apply any translations to the NRs, as randomly selected measured NRs should mimic the expected spatial distribution of NRs in the MIGDAL OTPC. Figure \ref{fig:hybrid} shows visual examples of hybrid Migdal construction.

\item \textbf{Hybrid coincidence events}: Using the same sample of around 27,000 NR frames from measurement, simulated ERs are redrawn at random and placed in a random location on the measured NR frame, ensuring that no ER pixel is within 20 pixels of any edge of the readout. After ERs are placed, gain and vignetting scalings are applied to the ER.

\end{enumerate}
All together we use around 27,000 frames of each of the two pure simulation classes and a similar number of frames of each of the two hybrid simulation classes.

In terms of evaluating YOLO's Migdal detection performance, one would ideally evaluate YOLO's performance on Migdal samples generated from both measured ERs and measured NRs; however this is not practical for the following reasons:

\begin{figure*}[htbp]
\centering
\includegraphics[width=0.8\textwidth]{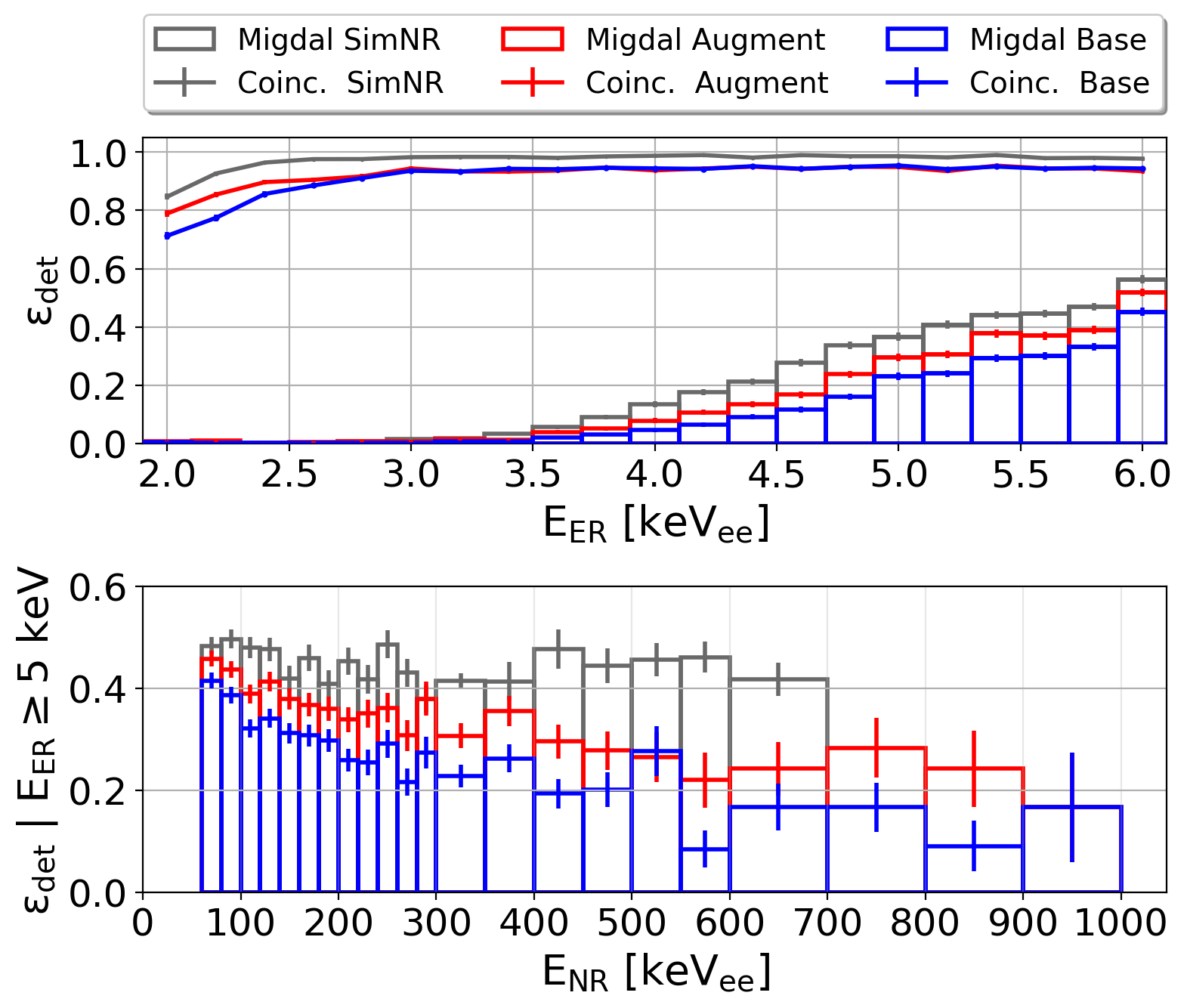}
\caption{Top: Simulated Migdal (bars) and coincidence (solid curves) detection efficiency as defined in Eq.$\,$(\ref{eq:3.4}) versus ER energy, integrated over all NR energies. Bottom: Migdal detection efficiency versus NR energy for pure simulated and hybrid Migdal events with $\SI{5.0}{keV_{ee}}\leq E_\mathrm{ER}\leq\SI{6.0}{keV_{ee}}$. Simulated NR energies are ground truth energies with SRIM quenching factors applied, which is why the pure simulation histogram terminates at a lower NR energy than the hybrid simulation histograms. The training and evaluation samples for each color plotted in this figure are summarized in Table$\,$\ref{tab:iii}.\label{fig:migeff}}
\end{figure*}

\begin{enumerate}
    \item Vertices of ERs are challenging to correctly locate. Unlike NRs, tails of ER tracks tend to be regions with relatively low light yield due to their characteristic Bragg curves. Furthermore, ERs tend to form tracks with more meandering paths than NRs, further complicating efforts to accurately detect measured ER vertices.
    \item Measured ERs also have noise, thus in regions where the ER is stitched to the NR frame, noise would be double counted, which could bias deep learning-based methods of track identification.
\end{enumerate}

There are clear advantages to considering hybrid simulation samples over pure simulation. Most notably, we observe larger halos forming around the outer edges of measured NR tracks that are not present in simulation$\,$\cite{Knights:2024wjh}. These halos are especially apparent in higher energy NRs and result in an increased area of observed NR tracks compared to their simulated counterparts, further obscuring Migdal electrons. Additionally, simulating Migdal events with measured NRs has the advantage of ensuring the correct gains, energy distributions, and spatial distributions of NRs when assessing YOLO's Migdal detection performance. To keep these performance studies general, all ERs considered in these studies have isotropic angular distributions. 

\subsection{Detection results: Simulated Migdal efficiencies versus energy}
\label{subsec:SimResults}

Quantifying Migdal detection efficiencies as a function of ER energy based on 2D track information from the OQC is an essential step toward deriving the total Migdal efficiency of the experiment, which will involve information from other detector subsystems such as the ITO and PMT. For each of the simulated Migdal and coincidence samples, we define YOLO's detection efficiency as
\begin{align}
    \label{eq:3.4}
    \mathrm{\varepsilon_{det}} \equiv \frac{N_\mathrm{det}}{N},
\end{align}
where $N$ is the total number of frames in the sample and $N_\mathrm{det}$ is the number of frames with a positive detection. A positive detection satisfies the following criteria: (1) YOLO identified exactly one ER and one NR, and (2) YOLO correctly localized its ER and NR bounding boxes in that the IoU overlap between the predicted ER and its ground truth counterpart and the IoU overlap between the predicted NR and its ground truth counterpart are both greater than zero. In other words, we require that $\mathrm{IoU}(B^\mathrm{ER}_\mathrm{p},B^\mathrm{ER}_\text{t}) > 0$ and $\mathrm{IoU}(B^\mathrm{NR}_\mathrm{p},B^\mathrm{NR}_\text{t}) > 0$.

Figure \ref{fig:migeff} summarizes YOLO's Migdal and coincidence detection efficiency performance versus energy. The legend of this figure shows results for two distinct YOLO training campaigns and different evaluation sets which are summarized in Table$\,$\ref{tab:iii}.
\begin{table}[htbp]
    \centering
    \caption{Companion table for Fig.$\,$\ref{fig:migeff} showing the training and evaluation samples associated with each color in the figure. The Base training sample includes only the real data shown in Table$\,$\ref{tab:i}. The Augment training sample includes all training data from the Base sample, as well as an additional 3,500 pure simulated Migdal events and 3,500 pure simulated coincidences.}
    \label{tab:iii}
    \begin{ruledtabular}
    \begin{tabular}{lcr}
    Legend color & Training sample & Evaluation sample \\ \hline
    Gray & Augment & Pure simulation \\ 
    Red & Augment & Hybrid \\ 
    Blue & Base & Hybrid \\
    \end{tabular}
    \end{ruledtabular}
\end{table}

The top panel of Fig.$\,$\ref{fig:migeff} shows $\rm \varepsilon_{det}$ versus ER energy over all NR energies in our sample, with bars representing Migdal detection efficiencies on the Migdal sample and solid curves representing coincidence detection efficiencies on the coincidence sample. We find that YOLO trained on the Augment sample and evaluated on hybrid simulation (red) improves on low-energy ER detection in the coincidence sample and also achieves significantly higher Migdal detection efficiencies compared to YOLO trained only on the Base sample (blue). As expected, the Migdal detection efficiencies are highest on the pure simulation sample, suggesting that less of the ER is, on average, obscured by the NR, since simulated NRs do not have the additional halo surrounding the track. This is further supported by the bottom panel of Fig.$\,$\ref{fig:migeff}, which shows Migdal detection efficiencies as a function of $E_\text{NR}$ for the subsets of pure and hybrid-simulated Migdal events with $\SI{5.0}{keV_{ee}}\leq E_\text{ER}\leq\SI{6.0}{keV_{ee}}$. The hybrid subset has roughly a factor of two decrease in $\rm\varepsilon_{det}$ between the lowest NR energy bins and higher energy ($\gtrsim\,$\SI{300}{keV_{ee}}) bins that is not present in the pure simulation sample. Evidently, the halos surrounding NR tracks are more extreme for higher energy NRs, further obscuring Migdal electrons. Besides the smaller halos present in low energy NRs, their topology also becomes more spherical, leading to $\rm\varepsilon_{det}$ becoming independent of the angle between the NR and ER track directions. This increase in the angular phase space for detection leads to the rise in $\rm\varepsilon_{det}$ at lower $E_\mathrm{NR}$ that is observed in both the hybrid and pure simulation datasets.

Integrating the hybrid Migdal distributions from the bottom panel of Fig.$\,$\ref{fig:migeff} over all NR energies, we find $\rm \varepsilon_{det}$ for the subset of hybrid Migdal events with ERs satisfying our ROI threshold ($E_\mathrm{ER}\in\{5.0,5.2,5.4,5.6,5.8,6.0\}$\,$\SI{}{keV_{ee}}$) to be 0.29 and 0.35 for the Base and Augment samples, respectively. While the expected Migdal ER emission probability falls off exponentially with increasing ER energy$\,$\cite{Cox:2022ekg}, the contribution to these efficiencies from ERs with $E_\text{ER}>\SI{6}{keV_{ee}}$ is not negligible. Since we expect Migdal detection efficiencies to continue to increase for $E_\text{ER}>\SI{6}{keV_{ee}}$, these quoted efficiencies are underestimates of the true Migdal detection efficiencies over our entire $\SI{5}{keV_{ee}}\leq E_\text{ER}\leq \SI{15}{keV_{ee}}$ ROI. More relevant, however, is the efficiency, $\rm\varepsilon_{det}$, integrated over the hybrid Migdal NR spectrum and shown as a function of $E_\text{ER}$ in the top panel of Fig.$\,$\ref{fig:migeff}, as measurements expressed in this way can be used to compare with theory. YOLO's ability to detect ERs in Migdal events was integral in deriving these ER energy-dependent efficiencies, so it is natural to consider more broadly how well YOLO performs this task. We address this next where we describe a method to quantify YOLO’s performance in detecting overlapping tracks.

\subsection{Detection results: YOLO performance observables}
\label{subsec:SimResults2}

To address YOLO's overlapping track identification performance more generally, we define a detectability criterion where we call a truth ER pixel in a simulated Migdal event significant if at least 1/3 of its intensity comes from the truth ER track.\footnote{To compute the fractional composition of intensity for a given pixel, we use the truth ER's intensity after applying gain scaling, vignetting scaling, and noise. We do not perform any processing to the NR intensities since they come from measurement.} We can then define $\rm n_{sigpix}$ as the number of significant pixels in an event. We define a related quantity, $\rm f_{sigpix}$, as the fraction of ground truth ER pixels in an event that are significant
\begin{align}
    \mathrm{f_{sigpix}}\equiv\frac{\mathrm{n_{sigpix}}}{\mathrm{n_{ERpix}}},
\end{align}
where $\rm n_{ERpix}$ is the number of ground truth ER pixels. For convenience, we summarize definitions of $\rm n_{sigpix}$, $\rm n_{ERpix}$, and $\rm f_{sigpix}$ in Table$\,$\ref{tab:iv}.

\begin{table}[t]
    \centering
    \caption{Description of quantities related to our detectability criterion.\label{tab:iv}}
    \begin{ruledtabular}
    \begin{tabular}{ll}
        Quantity & Definition \\ \hline
         & Number of truth ER pixels where at\\
        $\rm n_{sigpix}$ & least 1/3 of the total pixel intensity\\
         & comes from the ER. \\
        $\rm n_{ERpix}$ & Total number of truth ER pixels. \\ 
        $\rm f_{sigpix}$ & $\rm n_{sigpix}\,/\,n_{ERpix}$ \\
    \end{tabular}
    \end{ruledtabular}
\end{table}

Evaluating Migdal detection efficiency as a function of $\rm f_{sigpix}$ (or related quantities) is a more intrinsic measure of YOLO's performance. Our ROI for Migdal searches covers a broad dynamic range, with peak NR intensities orders of magnitude more intense than peak ER intensities, so in events where a Migdal ER is fully embedded within the NR penumbra, we expect $\rm n_{sigpix}$ to be nearly zero. While comparing YOLO's Migdal detection performance in pure simulation samples to hybrid simulation was useful in the previous study to show the adverse effects NR halos have on Migdal detection, the presence of halos should merely shift the distribution of $\rm f_{sigpix}$ closer to zero. We therefore opt to only consider hybrid simulation when evaluating $\rm\varepsilon_{det}$ versus $\rm f_{sigpix}$\footnote{Also note; variations in Migdal ER energies and emission angles with respect to NR directions will affect the distribution of $\rm f_{sigpix}$ but should not affect $\rm\varepsilon_{det}$ versus $\rm f_{sigpix}$.}. We evaluate this relationship using YOLO trained on the Augment training sample (Table$\,$\ref{tab:iii}), as YOLO trained on this sample outperformed YOLO trained on the Base sample in Fig.$\,$\ref{fig:migeff}.

\begin{figure}[b]
\centering
\includegraphics[width=0.48\textwidth]{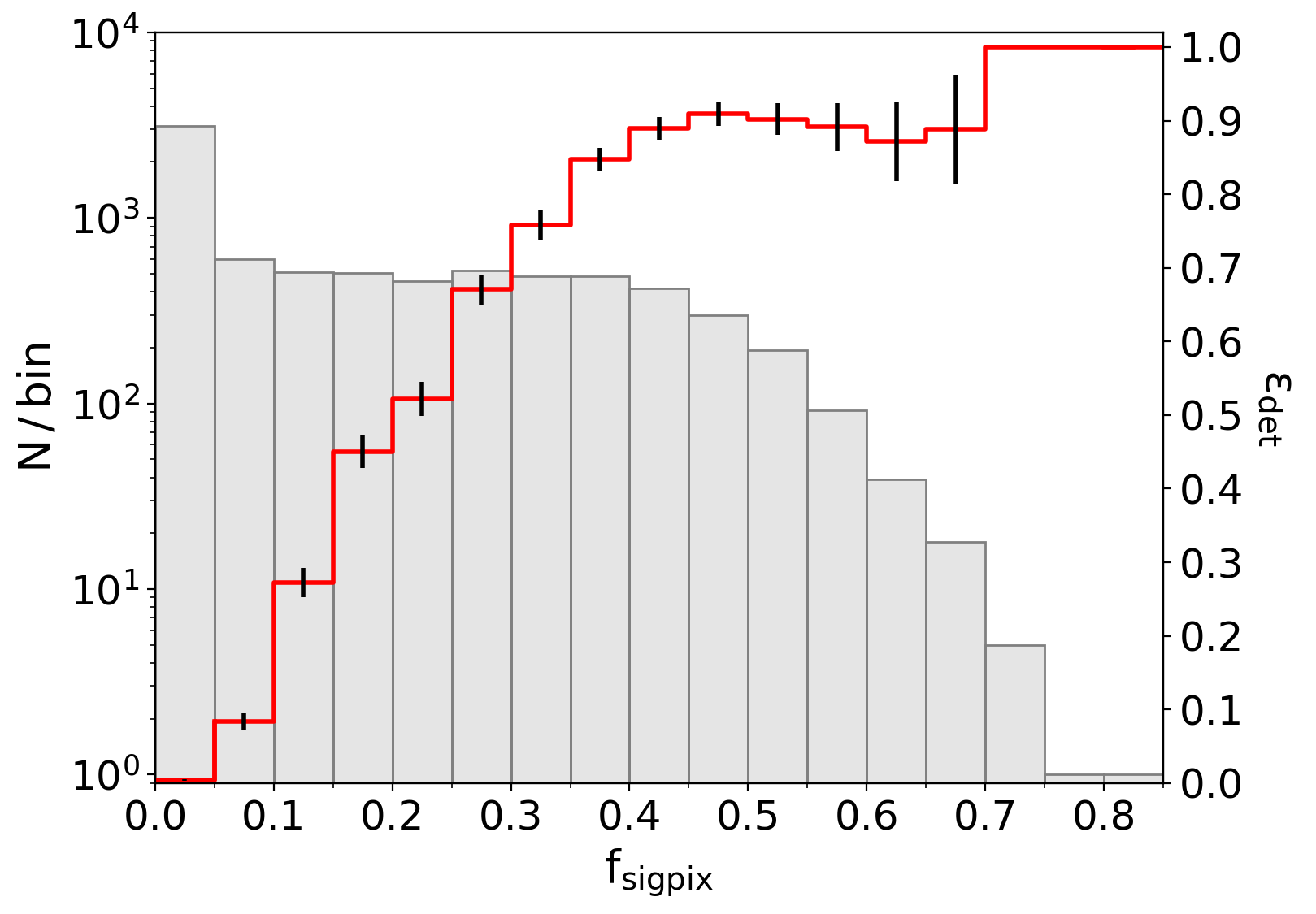}
\caption{Distribution of $\rm f_{sigpix}$ (gray bars; left vertical axis) and Migdal detection efficiency versus $\rm f_{sigpix}$ (red histogram; right vertical axis) of the $\SI{5.0}{keV_{ee}}\leq E_\text{ER}\leq\SI{6.0}{keV_{ee}}$-subset of the hybrid Migdal simulation set trained on the Augment training sample.\label{fig:migeffnum}}
\end{figure}

Figure \ref{fig:migeffnum} shows the results of this study on the $\SI{5.0}{keV_{ee}}\leq E_\text{ER}\leq\SI{6.0}{keV_{ee}}$-subset of this data. While we previously found a Migdal detection efficiency of 0.35 for all events in this sample, we see from this study that 40$\%$ of those events have $\rm f_{sigpix}<0.05$ and are therefore essentially undetectable ($\rm\varepsilon_{det}<0.01$). $\rm \varepsilon_{det}$ nearly doubles to 0.67 for the subset of this sample that also satisfies $\rm f_{sigpix}>0.1$, and further improves to 0.81 for $\rm f_{sigpix}>0.25$, demonstrating that YOLO exhibits excellent Migdal detection performance even in cases of significant ER-NR overlap, provided the ER is not completely enveloped by the NR. Mitigating NR track smearing through effects like optical halo formation, and, more broadly, reducing diffusion through the usage of a negative ion drift gas mixture should reduce the likelihood of a Migdal event having zero significant pixels and therefore significantly improve overall Migdal detection efficiencies.

\begin{figure}[t]
\centering
\includegraphics[width=0.45\textwidth]{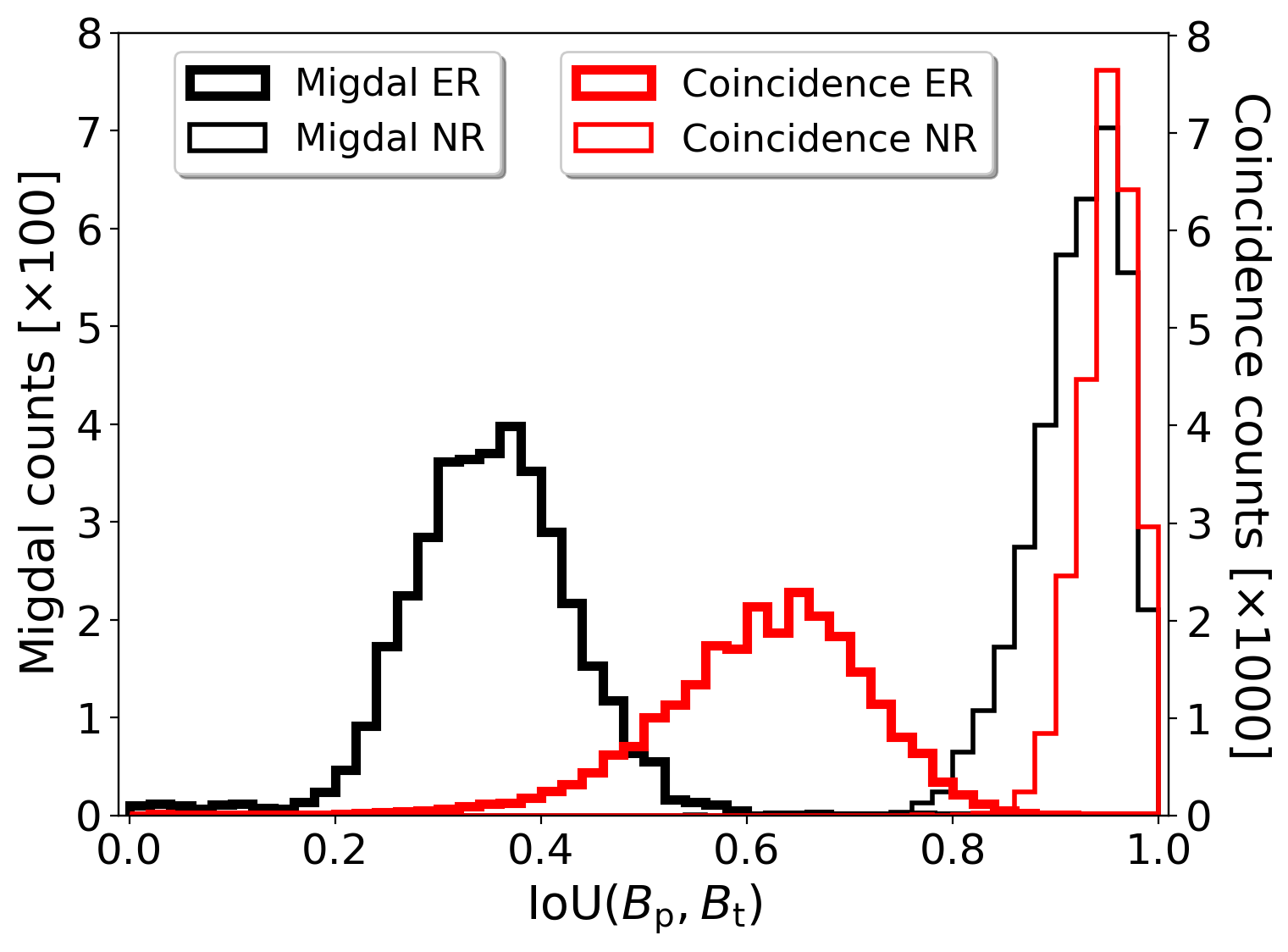}
\caption{Comparison of bounding box overlap between ground truth and YOLO's predictions for positive detections in the hybrid Migdal (black) and hybrid coincidence simulation (red) samples. The bolded curves represent ER IoU overlap [$\mathrm{IoU}(B^\mathrm{ER}_\mathrm{p},B^\mathrm{ER}_\mathrm{t})$], while the unbold curves represent NR IoU overlap [$\mathrm{IoU}(B^\mathrm{NR}_\mathrm{p},B^\mathrm{NR}_\mathrm{t})$]. \label{fig:IOU}}
\end{figure}

\subsection{Localization performance}

Our definition of a positive detection only requires the minimal localization requirement that $\mathrm{IoU}(B^\mathrm{NR}_\mathrm{p},B^\mathrm{NR}_\mathrm{t})$ and $\mathrm{IoU}(B^\mathrm{ER}_\mathrm{p},B^\mathrm{ER}_\mathrm{t})$ are each greater than zero. To more thoroughly quantify localization performance, we plot these IoU distributions for positive detections in both the hybrid Migdal and hybrid coincidence samples, which are shown in black and red in Fig.$\,$\ref{fig:IOU}, respectively. Overall, we observe excellent localization performance for NRs with mean IoUs between prediction and ground truth of 0.92 and 0.95 for positive detections in the hybrid Migdal and hybrid coincidence samples, respectively. ER localization performance is also very good, with a mean IoU of 0.61 for hybrid coincidences. Due to the larger variance in ER trajectory and lower pixel intensities compared to NRs, it is not surprising that ERs are not localized as well as NRs in general. The fact that there is a significant drop in the localization performance in the Migdal sample when compared to the coincidence sample is also not a surprise, as a significant fraction of ERs are obscured by the NR in the former sample.

\begin{figure}[t]
\centering
\includegraphics[width=0.45\textwidth]{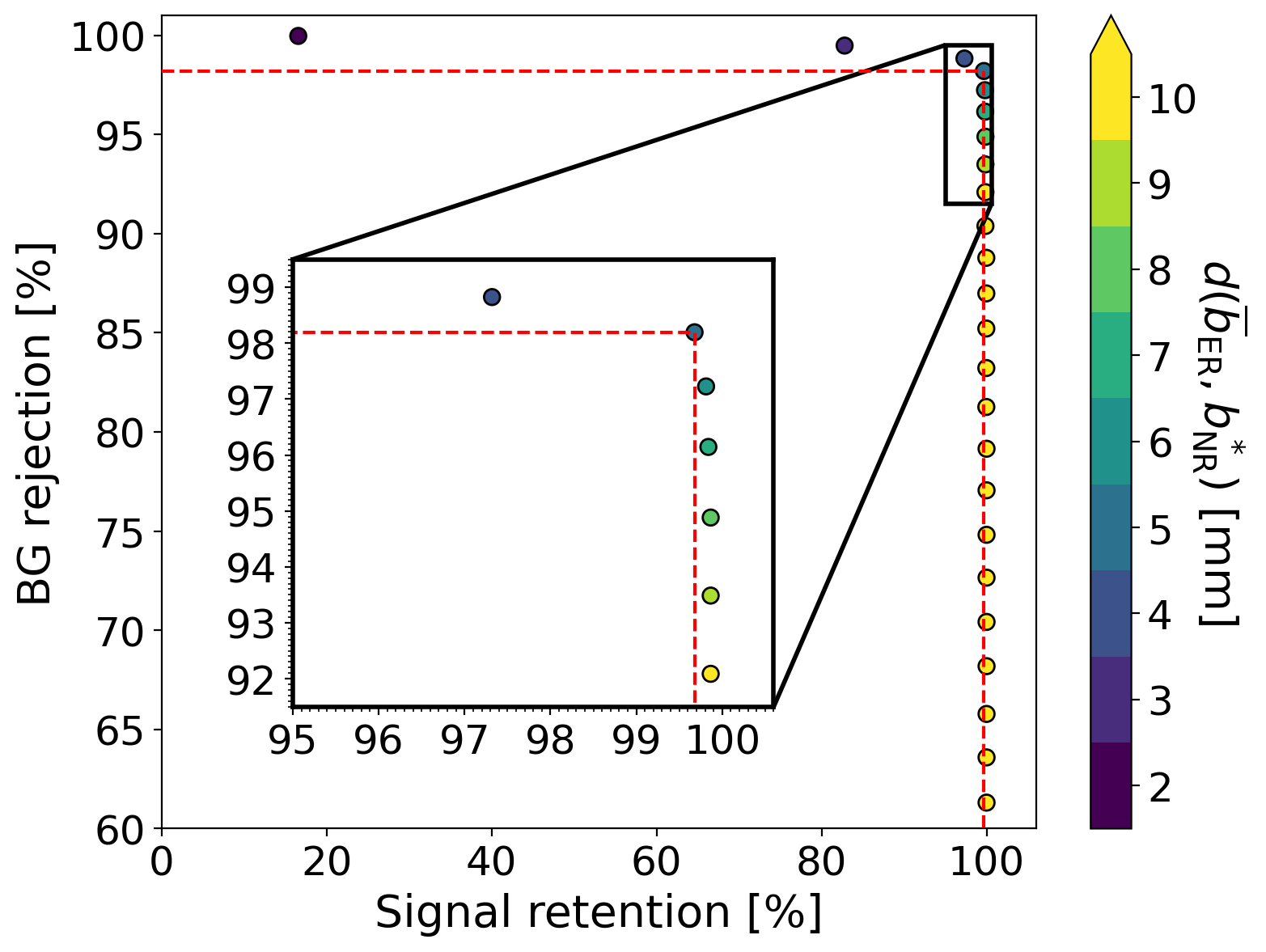}
\caption{Random coincidence background rejection and acceptance of simulated Migdal signal detected by YOLO shown at integral steps of $d(\overline{b}_\mathrm{ER},b^*_\mathrm{NR})$ -- the maximum allowed distance between the centroid of the ER bounding box and point of highest intensity of the NR bounding box. The zoomed-in inset shows the performance near the $d(\overline{b}_\mathrm{ER},b^*_\mathrm{NR})\leq\SI{5}{mm}$ optimum. Red dashed lines denote the background rejection and signal retention at this optimum.}
\label{fig:coincrej}
\end{figure}

\subsection{Background rejection study}
\label{subsec:coincrej}

We next perform an optimization study on coincidence background rejection and Migdal detection efficiency. Since ERs and NRs share a vertex in Migdal events, the distance between reconstructed ER and NR vertices would be the ideal background rejection metric. In practice, however, it is extremely challenging to accurately reconstruct the vertex position of keV-scale ERs. Our aim here is simply to characterize the background rejection potential of YOLO as a Migdal detection tool, so we will opt to use a simplified approximation of ER-NR vertex distance to do this. We therefore choose the separation distance, $d(\overline{b}_\mathrm{ER},b^*_\mathrm{NR})$, between the centroid of the ER bounding box, $\overline{b}_\mathrm{ER}$, and the point of highest intensity of the NR bounding box, $b^*_\mathrm{NR}$, as our metric to optimize background rejection. The background sample for this study consists of the roughly 25,000 coincidence frames with positive detections from YOLO trained on the Augment sample, while the signal sample is the subset of the hybrid Migdal sample with positive detections.


The results of this study are summarized in Fig.$\,$\ref{fig:coincrej} which shows the coincidence background rejection and signal retention at integral steps of $d(\overline{b}_\mathrm{ER},b^*_\mathrm{NR})$. We find $d(\overline{b}_\mathrm{ER},b^*_\mathrm{NR}) \leq \SI{5}{mm}$ maximizes the product of signal retention and background rejection, and choose this as our optimal point. Selecting frames satisfying $d\leq \SI{5}{mm}$ retains $99.7\%$ of detected Migdal events while rejecting $98.2\%$ of coincident backgrounds. Expressed in terms of raw counts, this selection reduces coincident background frames from 25,154 to 490, while retaining 3,719 of the 3,731 signal frames.

We have found the distance between the centroid of the ER bounding box and point of highest intensity of the NR bounding box to be an excellent coincidence background rejection discriminant. What remains to be tested is YOLO's false positive identification rate. We test this by evaluating our pipeline (with YOLO trained on the Augment sample) on a sample of 5,000 randomly selected frames containing a single simulated NR satisfying $E_\mathrm{NR}\geq\SI{60}{keV_{ee}}$ 
and nothing else. The simulated frames are processed following the procedure detailed in Appendix$\,$\ref{subsec:ERsim}. Of these 5,000 frames, YOLO falsely identifies 7 NRs as an ER-NR pair, with the remaining 4,993 correctly identified as a single NR, leading to a false positive rate of 0.14$\%$. While these false positive events could pass YOLO's initial Migdal search and be flagged as candidates in an online analysis, the combination of more detailed later stage offline image analyses and analyses with other subsystems should rule out these false positives.

\section{Real-time analysis and Migdal searches}
\label{sec:Reporting}

Our earlier benchmark showed that our pipeline is capable of achieving real-time speeds when performing an OQC end-to-end Migdal effect search on the MIGDAL readout PC. With this established, we begin this section by highlighting in more detail the online deliverables provided and displayed in real-time by our pipeline. We then conclude this section with the application of the steps of our online Migdal search to a very large sample of OQC data to illustrate the scale of the data reduction achieved when performing a Migdal effect search.

\subsection{Online deliverables}
\label{subsec:online}
Recall that our pipeline converts image-indexed data to track-indexed data using YOLO's bounding box assignments, allowing for efficient extraction of physics information on a track-by-track basis. For each track-index, we compute the following quantities online using the pixel content within the bounding box evaluated on $4\times 4$-binned, Gaussian smoothed images:
\begin{enumerate}
\item Intensity/energy: Sum up all pixels above threshold to compute the intensity of the track. We correct the intensity for vignetting (Appendix$\,$\ref{sec:A1}) and use appropriate $^{55}$Fe calibration-run data to convert the corrected intensity into energy in units of \SI{}{keV_{ee}}.
\item 2D length: Use a singular value decomposition$\,$\cite{1102314} to identify the track's principal axis in the OQC readout plane and compute the length of the track along that axis. Generally speaking, this method works better for NRs than ERs since NR topologies are typically better modeled by straight lines than ER topologies.
\item Head/tail identification (for ERs and NRs only): We split the track in half along its principal axis and count the intensity in each half of the track to identify the track's head and tail$\,$\cite{Hedges:2021dgz}. If YOLO identified the track as a NR (ER), we assign the side of the track with less intensity as the head (tail). Since this method is reliant on principal axis identification, it works best in cases where track trajectories are modeled well by straight lines.
\item 2D vector angle (ERs and NRs only): Use the assigned head/tail direction to compute the angle of the track's principal vector with the +$x$ axis.
\item 2D axial angle (protons and alphas only): Compute the angle between the track's principal axis with the +$x$ axis.
\end{enumerate}

\begin{figure*}[htbp]
\centering
\includegraphics[width=0.78\textwidth]{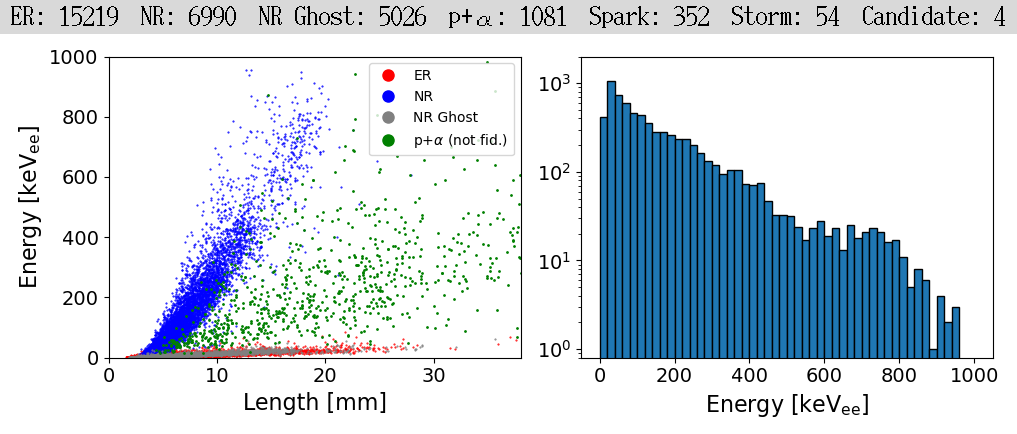}
\caption{Snapshot of a live online display that updates with 600 images worth of data every five seconds. The top banner shows updating counters of objects of interest accumulated over the course of the run. The left plot shows energy versus 2D length of tracks with color representing YOLO's classification assignments of track species of interest. The right plot shows the energy spectrum of NRs that YOLO identified. \label{fig:realtime}}
\end{figure*}

While we could substantially improve our 2D track fitting by training YOLO to identify key points along a track's trajectory (see Fig.$\,$\ref{fig:CV}), which would in turn improve head/tail identification, angular reconstruction performance, and likely still be implementable in real-time online analyses, this is beyond the scope of this work. Improving track fitting has no-bearing on the coincidence background rejection steps described in Sec.$\,$\ref{subsec:coincrej}, and is therefore not necessary for our pipeline's OQC-driven Migdal effect search, so we leave this for future work.

Some of the physical track quantities extracted by our pipeline are implemented into live online displays showing relevant event statistics during D-D runs. Figure \ref{fig:realtime} shows an example of such a display. The event counters and both plots refresh with new processed data from three image batches (600 frames) every five seconds. We operate this display on our readout PC during data acquisition, providing us with mixed-field particle identification, d$E$/d$x$ distributions of select classes of events, NR energy spectra, and event-class counts that include candidates satisfying user-input Migdal search criteria, all at real-time rates (\SI{120}{fps}) with at most a five second lag behind image acquisition. To calibrate NR energies online, we use the peak of a Gaussian fit to the most-recently recorded $^{55}$Fe spectrum (at the same nominal voltage across both GEMs). These $^{55}$Fe calibration fits are performed on online-processed data.

The count of four candidates shown in the banner of Fig.$\,$\ref{fig:realtime} represent ER-NR pairs satisfying Migdal search criteria of $E_\mathrm{NR}\geq\SI{60}{keV_{ee}}$ and $d(\overline{b}_\mathrm{ER},b^*_\mathrm{NR})\leq \SI{5}{mm}$. Recall that each time a candidate is recorded, a separate file containing the ER-NR pair-specific information and associated metadata is saved, allowing for immediate analysis of candidate-events.

In addition to performing Migdal candidate searches, our pipeline's online reporting is used alongside ITO detector data to provide immediate feedback on effective gains during $^{55}$Fe calibration runs. At a constant voltage across the GEMs, the presence of D-D neutrons causes a degradation in effective gain over the course of a day$\,$\cite{Knights:2024wjh}, so this feedback is crucial for making suitable adjustments to GEM voltages that ensure reasonably stable long-term gain. Online-processed data has also been useful for performing data-informed D-D generator and collimator alignments. All of these examples are made possible by the level of detail of the information extracted by our pipeline coupled with YOLO's real-time inference speed.

\subsection{An offline search for the Migdal effect}

\label{subsec:migskim}
\begin{table}[b]
\caption{\label{tab:v}%
Summary showing the number of camera frames remaining after increasingly restrictive selections throughout a Migdal candidate search over frames with \SI{8.33}{ms} exposure. Each row in the table incorporates the selections from all previous rows. Employing a \SI{60}{keV_{ee}} NR energy restriction and a \SI{5}{mm} maximum separation distance between the ER bounding box centroid and NR point of highest intensity reduces our sample from about 20 million frames to 826 frames.
}
\begin{ruledtabular}
\begin{tabular}{lr}
Selection & $N_\mathrm{frames}$\\
\hline
None (all frames analyzed by YOLO) & 19,996,200 \\
At least one fiducial ER-NR pair & 25,105 \\
$E_\mathrm{NR}\geq\SI{60}{keV_{ee}}$ & 15,121 \\
$d(\overline{b}_\mathrm{ER},b^*_\mathrm{NR})\leq\SI{5}{mm}$ & 826
\end{tabular}
\end{ruledtabular}
\end{table}

Our online analysis identifies candidates with 2D topologies that are consistent with the Migdal effect in real-time over the course of a run. It is therefore only necessary to perform an offline search for the Migdal effect if we want to integrate other detector subsystems with the optical system. That being said, it is illustrative to highlight the scale of the data reduction achieved when applying our online Migdal search criteria to a large sample of OQC images, so as an exercise, we apply these search criteria to an offline sample consisting of nearly 20 million images (\SI{90}{TB} raw, uncompressed images) recorded at the peak \SI{120}{fps} acquisition rate of the camera. Evaluating this sample using YOLO trained on the Base training set (Table$\,$\ref{tab:i}) and applying the search criteria of $E_\mathrm{NR}\geq\SI{60}{keV_{ee}}$ and $d(\overline{b}_\mathrm{ER},b^*_\mathrm{NR})\leq \SI{5}{mm}$ to frames containing at least one fiducial ER-NR pair (Table$\,$\ref{tab:v}) ultimately reduces the sample from 20 million frames to 826 frames.

\begin{figure*}[htbp]
\centering
\includegraphics[width=0.76\textwidth]{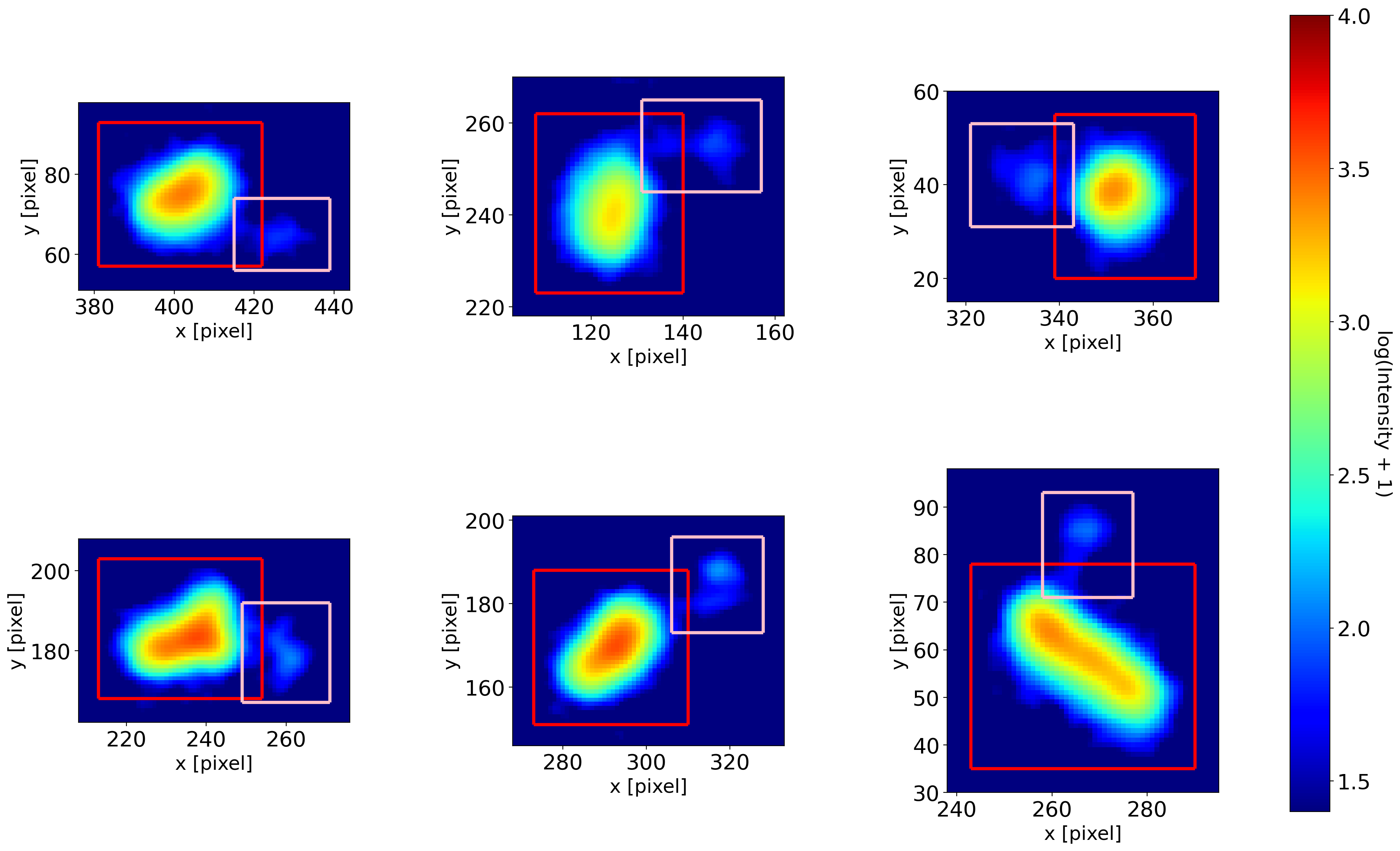}
\caption{Selection of six event displays satisfying all search criteria applied in Table$\,$\ref{tab:v}. The frames are $4\times 4$-binned with Gaussian smoothing applied, leading to a \SI{3.1}{mm} per twenty pixel conversion. NR bounding boxes are colored red and ER bounding boxes are colored pink. Clockwise starting from the upper left frame, the reconstructed energies $\{E_\mathrm{NR},E_\mathrm{ER}\}$, in units of \SI{}{keV_{ee}}, within YOLO's predicted bounding boxes are (i) $\{140,5.8\}$, (ii) $\{68,5.3\}$, (iii) $\{110,5.9\}$, (iv) $\{160,5.4\}$, (v) $\{170,5.3\}$, and (vi) $\{200,5.3\}$. \label{fig:migskim}}
\end{figure*}

Walking through Table$\,$\ref{tab:v} step-by-step, the selections listed in each row are applied successively. We first select frames containing at least one fiducial ER-NR pair satisfying (1) the ER is not flagged to be a NR ghost, (2) the NR is not clipped by the rolling shutter, and (3) the IoU overlap between the ER and NR is not greater than 0.95.\footnote{The $\rm IoU\leq 0.95$ restriction is to ensure that YOLO did not draw two nearly identical bounding boxes around the same track. This happens occasionally for low energy NRs where YOLO makes two plausible predictions; one being an ER and the other being a NR with nearly identical bounding boxes.} Applying this selection reduces our sample from 20 million frames to around 25,000 frames. In each of these 25,000 frames, we next link all unique ER-NR pairs together so we can conveniently make selections on an ER-NR track-pair basis. We then apply a \SI{60}{keV_{ee}} NR threshold to each track-pair, further reducing the sample to around 15,000 frames. In Sec.$\,$\ref{subsec:coincrej} we determined $d(\overline{b}_\mathrm{ER},b^*_\mathrm{NR})\leq\SI{5}{mm}$ to be the optimal selection for simultaneously reducing coincidence backgrounds with minimal effect on signal acceptance. Applying this criterion to the remaining 15,121 frames leaves us with 826 frames.

Figure \ref{fig:migskim} shows a selection of six frames satisfying all selections in Table$\,$\ref{tab:v}. The orientations and separations of the identified ERs and NRs in each of these events appear consistent with a 2D Migdal effect topology hypothesis. However, the OQC alone is not sufficient for confirming the Migdal effect, so these events need to be further analyzed with the remaining detector subsystems to test whether or not they are consistent with the Migdal effect in 3D. Still, this figure illustrates that applying our search criteria on real data extracts the classes of events we are interested in. Moreover, on simulation we found the $d(\overline{b}_\mathrm{ER},b^*_\mathrm{NR})\leq\SI{5}{mm}$ selection rejected 98.2$\%$ of coincidence backgrounds while retaining $99.7\%$ of the detected signal, making it unlikely that these selections rejected any Migdal candidates identified by YOLO in the original 20 million frames. Through this exercise, we have demonstrated that applying our online search criteria on a sample of 20 million frames reduces the sample to 826 frames with simulation suggesting essentially no loss of detectable signal. This reduction transforms our rare event search into a much more manageable search, enabling detailed analyses of 2D candidates in all detector subsystems.

\section{Broader Applications}
\label{sec:applications}
Using the MIGDAL experiment's search for the Migdal effect as an example, we have demonstrated object detection to be a favorable strategy for detecting composite rare event signals. By reframing our Migdal effect search as one for pairs of ERs and NRs in sufficiently close proximity, we were able to train YOLOv8 on an abundance of measured data (Table$\,$\ref{tab:i}), rather than exclusively on a simulated rare event signal. In Sec.$\,$\ref{subsec:SimResults2}, we quantified YOLO's detection performance of Migdal effect topologies constructed by stitching together simulated ERs with measured NRs, as a function of the fraction of truth ER pixels that are significant.\footnote{Those where at least 1/3 of the pixel intensity belongs to the truth ER; details in Sec.$\,$\ref{subsec:SimResults2}.} This study serves more broadly as useful proxy for assessing an object detection algorithm's performance in identifying composite events as a function of overlap. In cases with essentially complete overlap ($\rm f_{sigpix}<0.05$), Migdal ERs are essentially undetectable ($\rm \varepsilon_{det} < 0.01$) due to NR pixel intensities, on average, being orders of magnitude more intense than ER pixel intensities within our search ROI. In cases with more modest but still significant overlap, YOLO performs remarkably well (e.g. $\rm \varepsilon_{det} = 0.85$ for $\rm 0.35 \leq f_{sigpix} < 0.4$). For reference, the median total numbers of ER and NR pixels in an event in this sample are 610 and 1,694, respectively. Although we have not quantified our composite rare event signal detection approach outside of MIGDAL, it is useful to consider the adaptability of our approach to other applications.

Vertex identification and reconstruction is an essential component of many applications in nuclear and particle physics that can benefit from the explicit detection of multiple tracks sharing a common vertex. In searches for decay processes where the decay constituents are commonly observed signatures, an object detection algorithm like YOLO could be applied in the exact same manner as we have done with MIGDAL. In particular, the algorithm can be trained on real data to identify the common decay components in isolation, and then when the trained algorithm is evaluated on science data, identified bounding boxes can be used to detect potential events of interest and reject backgrounds. Further analysis on the contents of the bounding boxes of events of interest can then be used to reconstruct the full track topology. The benefits of using our approach for vertexing are threefold: (i) the object detection model can be trained on real data, thus avoiding Sim2Real gaps and ensuring model performance will generalize well to real data, (ii) object detection classifies and localizes all decay constituents allowing for a full event reconstructions even in cases when some are moderately to heavily obscured, and (iii) vertex reconstruction/identification can be tuned using physically motivated quantities, rather than esoteric score-based quantities output from a deep learning model trained directly to identify the (rare) decay.

It is important to emphasize here that use of object detection is not restricted to optical readout systems. In fact, any readout capable of producing data that can be expressed as 2D images\footnote{Object detection can also be applied to 3D images. Due to the sparsity of volumetric data, segmentation algorithms may be better suited for 3D data than object detection.} -- including, but not limited to, wire, strip, pad, and pixel readouts -- can use object detection.

Finally, we note that segmentation algorithms may have the potential to better characterize regions of heavy pixel overlap than object detection, making them very attractive candidates for vertex identification and reconstruction applications. We do not, however, consider segmentation algorithms here because more work is needed to explore how to best handle regions of pixel overlap when training on real data, as \textit{a priori} knowledge of a pixel's identity is not known in such regions. Even if sufficient information were known to train an algorithm to characterize regions of heavy pixel overlap, the labor required to hand-label segmentation boundaries in order to form a training set based on real-data, would be a monumental task.

We now consider some specific examples of where our approach of using object detection to reconstruct rare events could be adapted to other experiments.

\subsection{Neutrinoless double beta decay}

Next-generation neutrinoless double beta decay ($0\nu\beta\beta$) experiments aim to achieve up to $\mathcal{O}(10^{28}\,$yr$)$ half-life sensitivities, requiring fewer than around 0.1 background event per tonne per year in their energy ROI$\,$\cite{Dolinski:2019nrj}. To achieve such an ambitious sensitivity, some next-generation $0\nu\beta\beta$ experiments like NEXT$\,$\cite{NEXT:2012zwy} and PandaX-III$\,$\cite{Chen:2016qcd} employ high pressure gaseous-$^{136}$Xe TPCs, which are capable of track topology reconstructions. While all next-generation $0\nu\beta\beta$ experiments are extremely radiopure and have excellent energy resolution, topological reconstructions provide an additional source of background rejection and means of signal confirmation$\,$\cite{Dolinski:2019nrj}. In particular, the $0\nu\beta\beta$ signal consists of two electrons, each of energy $Q_{\beta\beta}/2$, sharing a vertex, while the dominant backgrounds are single photoelectrons with energy near $Q_{\beta\beta}$.\footnote{$Q_{\beta\beta}$ is \SI{2.458}{MeV} in $^{136}$Xe$\,$\cite{Redshaw:2007un}.} The energy deposition near the head portion of an electron track in gaseous xenon rises with $1/v^2$, leading to a ``blob" at the head of the track with significantly higher d$E$/d$x$ than the rest of the track$\,$\cite{10.3389/fphy.2019.00051}. $0\nu\beta\beta$ candidates will therefore form two-headed tracks with blobs at each end that can be topologically distinguished from photoelectron background tracks of the same energy that have a distinct head and tail, and thus only one blob.

Both NEXT and PandaX-III produce $x$-$y$, $x$-$z$, and $y$-$z$ projections of event topologies, so our object detection approach could, in principle, be readily applied to these experiments. In particular, the object detection algorithm could be trained on real data to identify single electrons, and then, in frames containing multiple identified electrons, potential double beta decay events can be flagged using the distance between the bounding boxes. If two electrons are found satisfying this or similar criteria, further analysis to reconstruct the full 3D topology could be performed to determine whether both electrons originated from a common vertex.

There are a number of ways that this methodology could be further refined for this use-case. For one, the object detection algorithm could be trained specifically to locate blobs at the end of electron tracks so their $d$E/$d$x could be immediately computed. Another possible refinement would be checking the connectivity between blobs to rule out events where two blobs are not contiguously connected by a track.

Using a calibration source, NEXT has already performed data-driven discrimination studies in their NEXT-White demonstrator$\,$\cite{NEXT:2018rgj}. Pair production events at the \SI{1.593}{MeV} double escape peak of $^{208}$Tl have two blobs connected by a long and narrow track, and are therefore of similar structure to double beta decay events. Treating these pair production events as signal, NEXT quantified the discrimination between these events and backgrounds near this energy$\,$\cite{NEXT:2019gtz,NEXT:2020jmz,NEXT:2021vzd}. Of these studies, their highest reported background rejection was around $96\%$ at around $57\%$ signal acceptance using a Richardson-Lucy deconvolution-based$\,$\cite{Richardson:72} method that was later applied to their $0\nu\beta\beta$-search demonstration with NEXT-White$\,$\cite{NEXT:2023daz}. Similar to how we trained YOLO on 4$\times$4-binned, Gaussian smoothed images, an object detection algorithm could also be trained on the Richardson-Lucy deconvolved images that NEXT records, potentially leading to improvements in event selection. We remind the reader that in MIGDAL, we achieve greater than $80\%$ detection efficiency for faint low energy ERs only a few millimeters long, with up to $75\%$ of their truth pixels heavily obscured by overlapping bright, high energy NRs. We therefore expect the detection of such ERs in MIGDAL to be more challenging than detecting the double beta decay topological signature consisting of $\mathcal{O}$(\SI{10}{cm}) electrons of similar brightness, sharing a common vertex with significantly less spatial overlap.

\subsection{Exotic nuclear decays}
Optical gas TPCs are also used to study and directly image nuclear decays$\,$\cite{2007_Miernik}, including exotic charged decays of neutron-deficient isotopes such as $^{45}$Fe, $^{46}$Fe, $^{44}$Cr, and $^{48}$Ni$\,$\cite{miernik1,PhysRevC.90.014311}. All of the decays studied in these references involve the emission of some number of $\beta$'s, $\alpha$'s, and/or protons, so an object detection algorithm could be trained in a data-driven way on each of these decay species at desired energy ranges, as well as on the implanted ions. The algorithm could then immediately determine the number of decay species, classify them, and reconstruct each of their energies. The decay vertex could be reconstructed either using the overlap regions of all identified bounding boxes, or by using object-key point detection (Fig.$\,$\ref{fig:CV}) trained to reconstruct the trajectory of each identified object. If the latter approach is taken, the angles between decay species could be readily determined, with the vertex better localized than using object detection alone. Given the event rates and camera resolutions in these experiments, a fast object detection algorithm like YOLOv8 could feasibly be implemented for real-time decay detection, as we have demonstrated in MIGDAL.

\section{Summary}
\label{sec:summary}

We have developed a fast deep learning-based object detection pipeline that is trained on real-data, and applied to the MIGDAL experiment's rare event search for the Migdal effect. Training with a combination of real data and simulation and evaluating YOLO's performance on simulated Migdal effect events formed by stitching together measured NRs with simulated ERs, we derived a detection efficiency of $35\%$ for events with ER energies between \SI{5}{keV} and \SI{6}{keV}. While this efficiency appears low, in around $40\%$ of these events, fewer than $5\%$ of truth ER pixels are significant (Fig.$\,$\ref{fig:migeffnum}), so the ER track is essentially entirely obscured by the NR and therefore nearly impossible to detect. Requiring at least $10\%$ of truth ER pixels to be significant, the Migdal detection efficiency nearly doubles to $67\%$, and further improves to 81$\%$ in the sample where least $25\%$ of truth ER pixels are significant. These results quantify the degradation of Migdal detection efficiency with ER-pixel obscuration, signifying the importance of mitigating optical halo formation around NRs in our detector, as well as exploring the use of negative ion drift gas mixtures to reduce diffusion.

Using the MIGDAL readout PC, we benchmarked the pipeline's end-to-end processing and analysis speeds from raw image acquisition to Migdal candidate selection. The results showed that our pipeline consistently performs faster than the peak \SI{120}{fps} acquisition rate of the ORCA-Quest camera, and is capable of event processing rates near \SI{450}{Hz}. These speeds enable Migdal effect searches online, and in real-time. 

Migdal effect searches with our pipeline also exhibit excellent background rejection. Accidental coincidences, where independent ERs and NRs that are spatially coincident during the exposure time of an image frame, are the dominant source of ER-NR pairs occupying the same frame. These are the largest background for Migdal effect searches when using only the Orca-Quest readout. Setting the Euclidean distance between the centroid of the ER bounding box and the point of highest intensity of the NR bounding box to be less than or equal to \SI{5}{mm}, we rejected 98.2$\%$ of these backgrounds while retaining 99.7$\%$ of the simulated Migdal events detected by YOLO. We applied this selection and a \SI{60}{keV_{ee}} NR energy threshold to a sample of 20 million Orca-Quest camera frames collected by the MIGDAL experiment, and found only 826 frames passing these criteria, thereby transforming our rare event search into a not-so-rare event search.

Our work introduces the use of object detection as an approach to reconstruct composite rare event signals via their constituent parts, opening up the possibility of data-driven ML training in cases where the rare event signal is composed of common signatures. This approach should be of broad interest to other rare event search experiments, as it alleviates the need to train ML algorithms on simulated rare event signals in order to reap the benefits of a ML-driven search.

\section*{Data Availability Statement}
Data supporting this study are openly available at the GitHub and Zenodo repositories at$\,$\cite{schueler_2024_12628437}.

\begin{acknowledgments}
This work has been supported by the UKRI’s Science $\&\,$Technology Facilities Council through the Xenon Futures R$\,\&\,$D programme (awards ST/T005823/1, ST/T005882/1, ST/V001833/1, ST/V001876/1), Consolidated Grants (ST/W000636/1, ST/X006042/1, ST/T000759/1, ST/W000652/1, ST/S000860/1, ST/X005976/1), and TM’s and LM’s PhD scholarships (ST/T505894/1, ST/X508913/1); by the U.S. Department of Energy, Office of Science, Office of High Energy Physics, under Award Number DE-SC0022357; by the U.S. National Science Foundation under Award number 2209307; ET acknowledges the Graduate Instrumentation Research Award funded by the U.S. Department of Energy, Office of Science, Office of High Energy Physics; by the Portuguese Foundation for Science and Technology (FCT) under award number PTDC/FIS-PAR/2831/2020; and by the European Union’s Horizon 2020 research and innovation programme under the Marie Skłodowska-Curie grant agreement No. 841261 (DarkSphere) and No. 101026519 (GaGARin). ELA acknowledges the support from Spanish grant CA3/RSUE/2021-00827, funded by Ministerio de Universidades, Plan de Recuperacion, Transformacion y Resiliencia, and Universidad Autonoma de Madrid. KN acknowledges support by the Deutsche Forschungsgemeinschaft (DFG, German Research Foundation) under Germany's Excellence Strategy -- EXC 2121 "Quantum Universe“ -- 390833306. We are grateful to the Particle Physics Department at RAL for significant additional support which made this project possible. Thanks are also due to the CERN RD51 collaboration for their support through Common Project funds, hardware tests and training, and useful discussions. We would also like to thank the ISIS facility for technical assistance and for hosting this experiment. We thank Master’s students M.~Handley (Cambridge) and R.~Hafeji (Surrey).

For the purpose of open access, the authors have applied a Creative Commons Attribution (CC BY) license to any Author Accepted Manuscript version arising from this submission.
\end{acknowledgments}

\appendix

\section{Single track simulation}
\label{sec:AppendixSingleTrack}
The results highlighted in Sec.$\,$\ref{sec:Performance} rely on realistic optical image simulations of ERs and NRs. Here we validate our optical simulation procedure and quantify YOLO's\footnote{In this appendix, we use YOLO trained only on the Base, real data sample (See Sec.$\,$\ref{subsec:SimResults}).} detection performance on single tracks. In Appendix$\,$\ref{subsec:ERsim}, we detail the process of producing realistic optical image simulation using simulated $^{55}$Fe x-ray tracks. We start with $^{55}$Fe both because its spectrum covers a relatively small dynamic range, and because it allows us to tune and compare effective gains in simulation and measurement. Beginning with the procedure described in Ref.$\,$\cite{Araujo:2022wjh} to produce digitized optical pixel intensities of $^{55}$Fe tracks, we apply additional intensity scalings and position-dependent vignetting, and finally add noise from randomly selected dark frames. We verify both this procedure and YOLO's ER identification performance by comparing the simulated $^{55}$Fe intensity spectrum to an $^{55}$Fe intensity spectrum reconstructed from a measured sample of $^{55}$Fe x-rays in the presence of ERs from the D-D generator. Next, Sec.$\,$\ref{subsec:single_trackSim} studies YOLO's ER and NR track-identification performance more broadly, using frames containing only single tracks. These frames contain either simulated ERs drawn from a discrete uniform energy spectrum or simulated NRs drawn from a continuous energy spectrum.

\begin{table}[b]
\caption{\label{tab:ii}%
Electron identification performance summary on simulated frames containing a single $^{55}$Fe track. Columns from left to right: (1) Truth decay mode, (2) total number of frames where YOLO identified a single $\rm ^{55}Fe$ track (true positive detections), (3) number of frames where YOLO identified $>$1 $\rm ^{55}Fe$ track, (4) number of frames where YOLO did not identify any tracks, and (5) total number of frames. Frames where YOLO did not identify any tracks mostly consist of tracks located near the edge of the readout, so vignetting heavily suppresses their intensities to below threshold. On the other hand, most instances of YOLO identifying $>$1 ER bounding box were cases where YOLO drew multiple boxes around the same track. In a few instances, YOLO drew a bounding box around noise.
}
\begin{ruledtabular}
\begin{tabular}{lcccr}
Mode & 1 ER & $>$1 ER & No tracks & Total\\\hline
K-$\alpha$ & 60,920 (98.1$\%$) & 456 (0.7$\%$) & 720 (1.2$\%$) & 62,096 \\
K-$\beta$ & 9,113 (97.6$\%$) & 115 (1.2$\%$) & 108 (1.2$\%$) & 9,336
\end{tabular}
\end{ruledtabular}
\end{table}

\subsection{Simulated $^{55}$Fe generation and identification}
\label{subsec:ERsim}
Adequate low-energy electron identification (EID) and localization is essential for detecting Migdal effect candidates. To assess YOLO's EID performance and validate our simulation of detector effects, we begin by evaluating YOLO on simulated $^{55}$Fe tracks. Since this version of YOLO was trained exclusively on measurement, and there are inevitable differences between simulated and observed tracks, we expect the results shown here to underestimate the EID performance of YOLO when evaluated on measurement.

We generate 71,432 frames, each containing a single $^{55}$Fe track from either the \SI{5.9}{keV} K-$\alpha$ or \SI{6.5}{keV} K-$\beta$ decay mode (statistical breakdown in Table$\,$\ref{tab:ii}) following the general optical readout simulation procedure described in Ref.$\,$\cite{Araujo:2022wjh}, but binning to the dimensions of the OQC. We then perform the following procedure to incorporate further effects in the optical system:

\begin{enumerate}
\item \textbf{Perform gain scaling:} Scale simulated tracks' pixel intensities using an empirically determined multiplicative factor with resolution smearing.
\item \textbf{Apply vignetting scaling:} Vignetting in the optical readout causes pixel intensities to decrease with distance from the center of the readout. We simulate this effect by first randomizing the location of simulated tracks along the readout (shifting using random uniform distributions in $x$ and $y$) and then applying the following correction to each pixel in the track
\begin{align}
I(x,y) = I_0(x,y)\frac{(a-r)^2}{a^2}.
\end{align}
Here, $I_0(x,y)$ is the gain-scaled pixel intensity of the pixel at location $(x,y)$, $r$ is the distance between the pixel at $(x,y)$ and the center pixel of the readout $(\overline{x},\overline{y})$, and $I(x,y)$ is the resulting intensity of the pixel at $(x,y)$ after applying vignetting scaling. We empirically assign $a$ to be \SI{95}{mm} to achieve good agreement between this model of vignetting and the vignetting we observe in a typical $^{55}$Fe run (see Appendix$\,$\ref{sec:A1}).
\item \textbf{Apply noise:} Randomly select a measured dark-subtracted dark frame from a sample of 800 such frames and add it to the simulated $^{55}$Fe frame.
\end{enumerate}

\begin{figure}[t]
\centering
\includegraphics[width=0.48\textwidth]{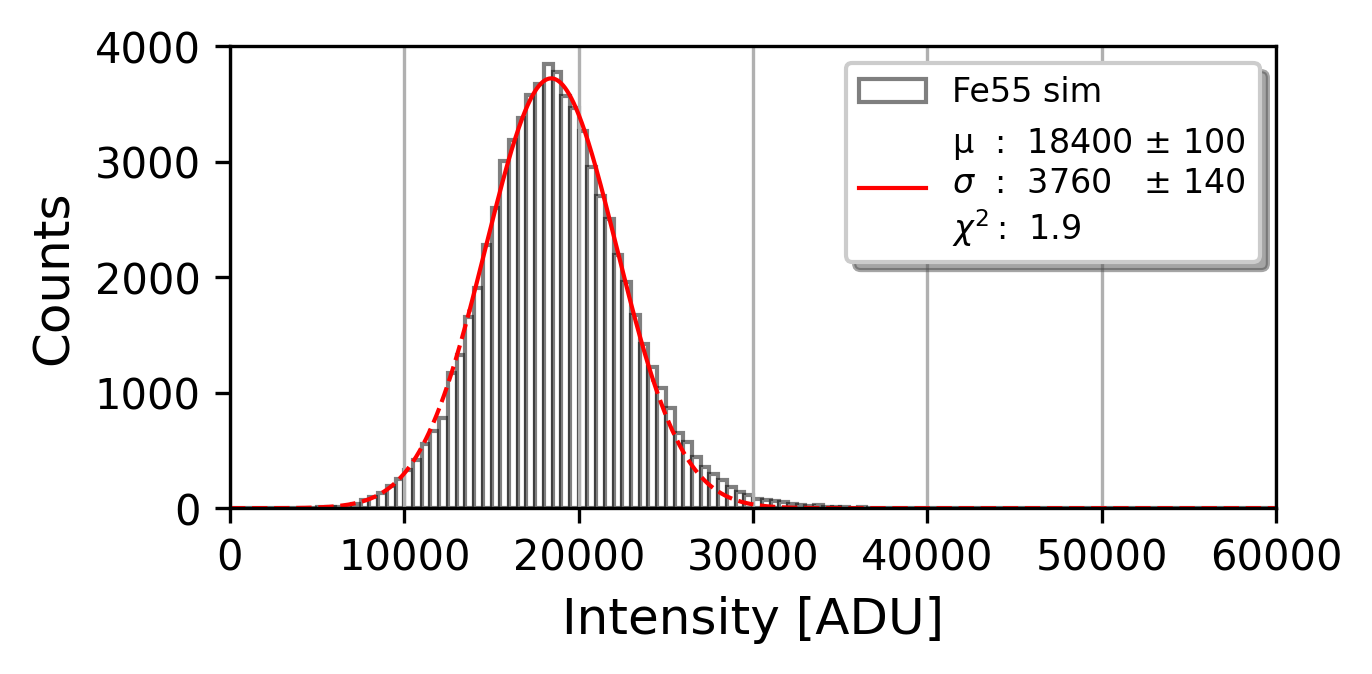}
\caption{Intensity spectrum of simulated $^{55}$Fe tracks identified by YOLO after applying gain scaling, vignetting, and adding noise. Data-driven vignetting corrections are computed following the procedure outlined in Appendix$\,$\ref{sec:A1} to achieve the shown spectrum. The solid line-portion of the curve fit is our fit-region for a single-peak Gaussian.\label{fig:simFe55}}
\end{figure}

Performing all steps of our pipeline on the resulting images, including vignetting corrections  (Appendix$\,$\ref{sec:A1}), we obtain the intensity spectrum shown in Fig.$\,$\ref{fig:simFe55}. While contributions of K-$\beta$ x-rays are not completely negligible, we opt to perform a single-peak Gaussian fit, so our reported resolution of $\sigma/\mu = 20.4\%$ is a slight overestimate of the true resolution of the \SI{5.9}{keV} ER peak. Table$\,$\ref{tab:ii} shows that YOLO trained on measurement is excellent at detecting simulated $^{55}$Fe tracks, with YOLO correctly predicting one and only one ER bounding box, $B_\mathrm{p}^\mathrm{ER}$, that has nonzero overlap with the ground truth bounding box of the frame ($\mathrm{IoU}(B_\mathrm{p}^\mathrm{ER},B_\mathrm{t}^\mathrm{ER})>0$) in 98.1$\%$ of K-$\alpha$ frames and 97.6$\%$ of K-$\beta$ frames. We note that the difference in false positive ER identification rates between the K-$\beta$ and K-$\alpha$ samples is small but significant. Further study is required to understand this difference, but we hypothesize the difference to be the result of K-$\beta$ tracks having slightly longer low-ionization tails on average than K-$\alpha$ tracks. If portions of these tails fall below the intensity threshold of the PNG images passed into YOLO, then YOLO may be more likely to assign multiple bounding boxes to the track. We expect training on a larger sample of higher energy ERs would resolve this discrepancy.

\begin{figure}[t]
\centering
\includegraphics[width=0.48\textwidth]{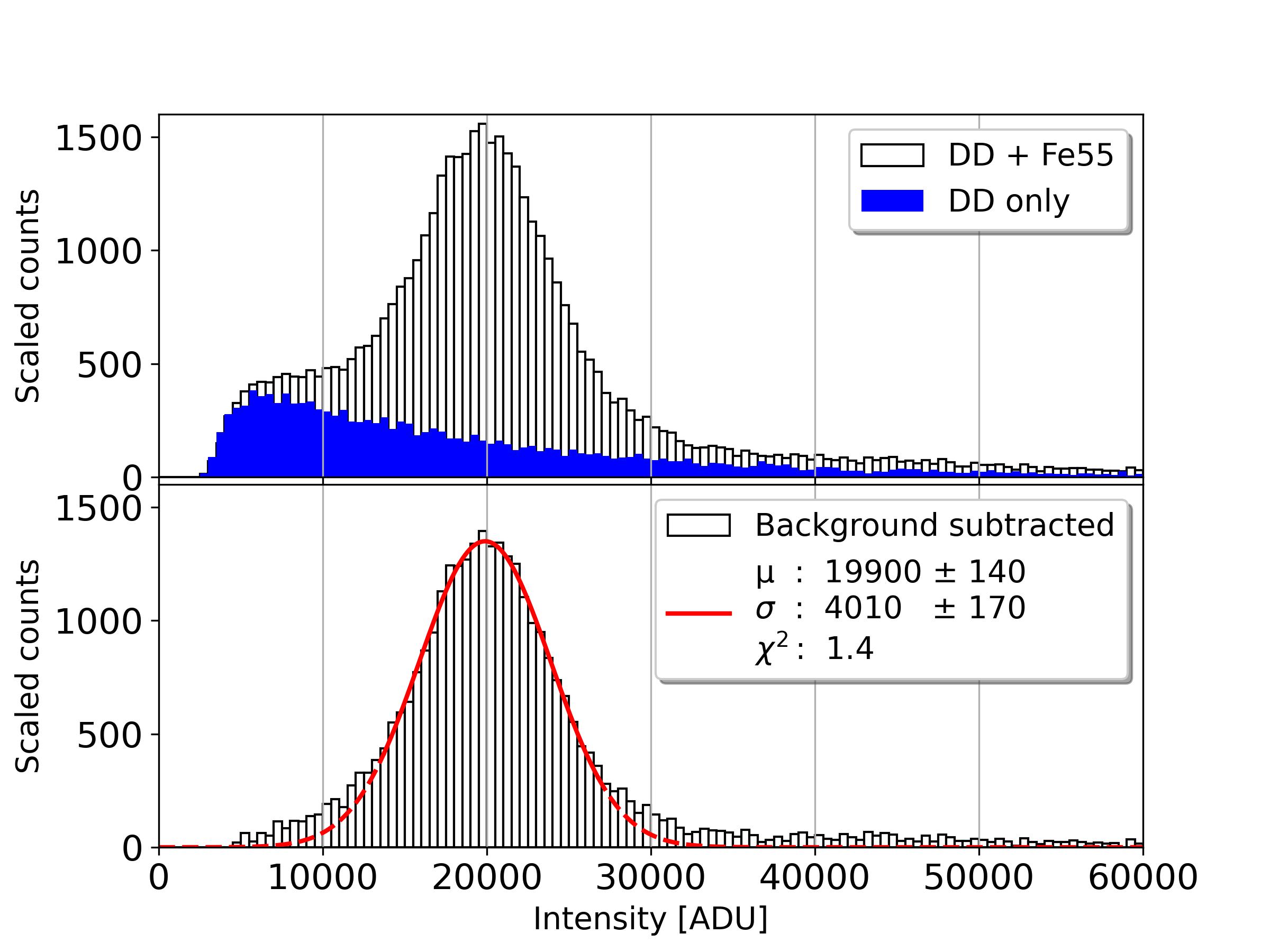}
\caption{Top: Electron recoil intensity spectrum for a D-D run with the $^{55}$Fe source present (``DD$\rm+^{55}$Fe"; black bars) and a D-D run without the $^{55}$Fe source present (``DD only"; blue shaded) scaled to the equivalent elapsed time of the D-D$\rm+^{55}$Fe run. Bottom: Recovered $^{55}$Fe spectrum after background subtracting the timescaled D-D run spectrum from the D-D$\rm+^{55}$Fe signal spectrum. The solid line-portion of the curve fit is our fit-region for a single-peak Gaussian.\label{fig:bg_subtract}}
\end{figure}

We next compare the simulated $^{55}$Fe spectrum fit in Fig.$\,$\ref{fig:simFe55} to a reconstructed $^{55}$Fe spectrum on data recorded in the presence of both the D-D generator and an $^{55}$Fe x-ray source. Performing such a comparison allows us to simultaneously validate our simulation of detector effects while qualitatively validating YOLO's EID performance on measurement through recovering the $^{55}$Fe peak in a mixed-field sample of data with a continuous background ER spectrum. Figure \ref{fig:bg_subtract} shows the results of this study with the top panel showing the ER intensity spectrum during the run with both the D-D-generator and $^{55}$Fe source present (labeled in legend as ``DD$+^{55}$Fe"), as well as the ``background" spectrum consisting of ERs recorded with only the D-D-generator (``DD only" in legend). NR ghosts are rejected from both spectra and the background spectrum counts are normalized to the elapsed time of the D-D$+\rm^{55}$Fe run. Subtracting this normalized background spectrum from the signal D-D$+^{55}$Fe data, we obtain the resulting $^{55}$Fe spectrum in the bottom panel of the figure. As with Fig.$\,$\ref{fig:simFe55}, we fit a single-peak Gaussian to this spectrum, assuming the peak at \SI{5.9}{keV}. We observe a peak of 19,900$\,$ADU and $20.2\%$ energy resolution, both of which are consistent with simulation when accounting for the $\sim$15$\%$ systematic gain variation over the course of an operating day in the presence of neutrons from the D-D generator (details Ref.$\,$\cite{Knights:2024wjh}), indicating that our simulation procedure properly models gain and $^{55}$Fe track light yield.

\subsection{General ER and NR identification on simulation}
\label{subsec:single_trackSim}
Next, we evaluate YOLO's ER and NR identification performance on frames containing a single simulated ER or NR of varying energy. We assess performance using localization efficiency, $\rm \varepsilon_{local}$, and detection efficiency, $\rm \varepsilon_{det}$ defined as
\begin{align}
    \label{eq:3}
    \mathrm {\varepsilon_{local}}&\equiv \frac{N(\mathrm{IoU}(B_\mathrm{p},B_\mathrm{t}) > 0)}{N} \\
    \label{eq:4}
    \mathrm {\varepsilon_{det}}&\equiv \frac{N\left((\mathrm{IoU}(B_\mathrm{p},B_\mathrm{t}) > 0) \land (y_\mathrm{p} = y_\mathrm{t})\right)}{N},
\end{align}
where $\land$ is the logical ``and" symbol, and $B_\mathrm{p}$ and $B_\mathrm{t}$ are YOLO's bounding box prediction and the ground truth bounding box with associated classifications $y_\mathrm{p}$ and $y_\mathrm{t}$, respectively. The numerator of Eq.$\,$(\ref{eq:3}) is the total number of frames where YOLO identified exactly one bounding box, $B_\text{p}$, and $B_\text{p}$ has nontrivial overlap with $B_\text{t}$. The numerator of Eq.$\,$(\ref{eq:4}) includes the additional restriction that the prediction associated with the bounding box agrees with the ground truth class of the event. The denominator of both of these equations is the total number of frames in the sample.

\begin{figure}[t]
\centering
\includegraphics[width=0.45\textwidth]{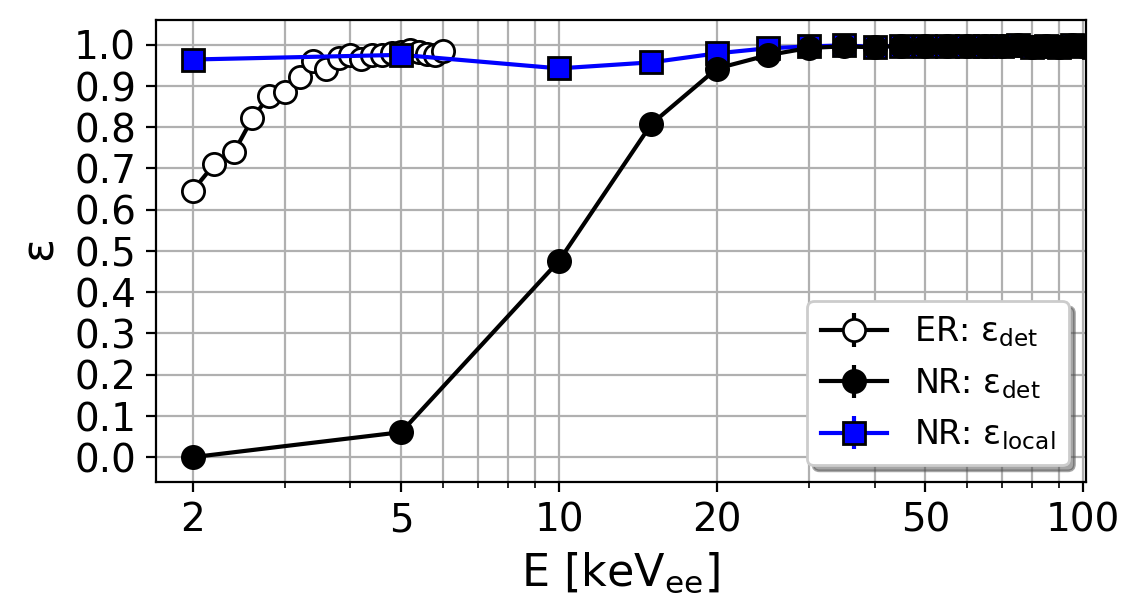}
\caption{Localization and detection efficiencies versus truth energy (SRIM quenching factors applied to NR energies) for frames containing single simulated tracks. For ERs, $\mathrm {\varepsilon_{det}}=\mathrm {\varepsilon_{local}}$, so we do not include a separate $\rm \varepsilon_{local}$ trace for ERs. \label{fig:YOLO_sim}}
\end{figure}

We evaluate $\mathrm {\varepsilon_{local}}$ and $\mathrm {\varepsilon_{det}}$ using the same sample of 10,500 simulated ERs described in Sec.$\,$\ref{subsec:twotracksim}, and 46,585 simulated NRs (36,779 fluorine recoils and 9,806 carbon recoils) with a continuous energy spectrum that mimics the expected energy distribution of D-D-induced nuclear recoils in the MIGDAL detector. Figure \ref{fig:YOLO_sim} shows YOLO's identification performance for all ERs and for NRs up to \SI{100}{keV_{ee}}. Energies in this figure are reported in \SI{}{keV_{ee}}, where SRIM$\,$\cite{SRIM} quenching factors have been applied to the ground truth simulated NR energies. Additionally, we find that for the truth ER samples, all bounding boxes identified are also predicted to be ERs, so $\mathrm {\varepsilon_{det}} = \mathrm {\varepsilon_{local}}$ for ERs. Given this, we only plot a single ER trace for ERs in the figure. Comparing the three traces, we make the following observations:

\begin{enumerate}
\item YOLO's detection efficiency exceeds $90\%$ for ERs down to \SI{3.2}{keV_{ee}} -- well below our \SI{5.0}{keV_{ee}} threshold -- and $70\%$ down to \SI{2.2}{keV_{ee}}.
\item The localization efficiency of \SI{2}{keV_{ee}} NRs far exceeds that of \SI{2}{keV_{ee}} ERs. This is likely due to the comparatively higher d$E$/d$x$ near the center of NR tracks, leading to regions of NRs with higher signal to noise ratios than ERs of comparable track energy.
\item Below \SI{10}{keV_{ee}}, YOLO misidentifies most truth NRs as ERs. Although troubling at first glance, we emphasize that our aim here is to efficiently select Migdal candidates in the 2D OQC while maximizing background rejection, so that identified candidates can be analyzed in all detector subsystems. There is a small probability of recording OQC frames where two independent NRs spatially coincide in 2D with one falling in the NR ROI energy range ($E_\text{NR}\geq\SI{60}{keV_{ee}}$) and the other in the ER ROI energy range ($\SI{5}{keV_{ee}}\leq E_\text{ER}\leq \SI{15}{keV_{ee}}$) for Migdal searches, resulting in a potential false positive in such a frame. Full 3D analyses of backgrounds in Ref.$\,$\cite{Araujo:2022wjh}, however, show that roughly one of these false positives is expected to occur for every 30 Migdal events. Such false positives are therefore not an issue when including information from other detector subsystems to reconstruct events in 3D. For this reason, the preponderance of misidentifying true NRs as ERs is actually preferential for our analysis, as we expect essentially zero false negative ER identifications, thereby maximizing ER detection efficiencies when searching for Migdal-like topologies.
\end{enumerate}

\begin{figure*}[htbp]
\centering
\includegraphics[width=0.84\textwidth]{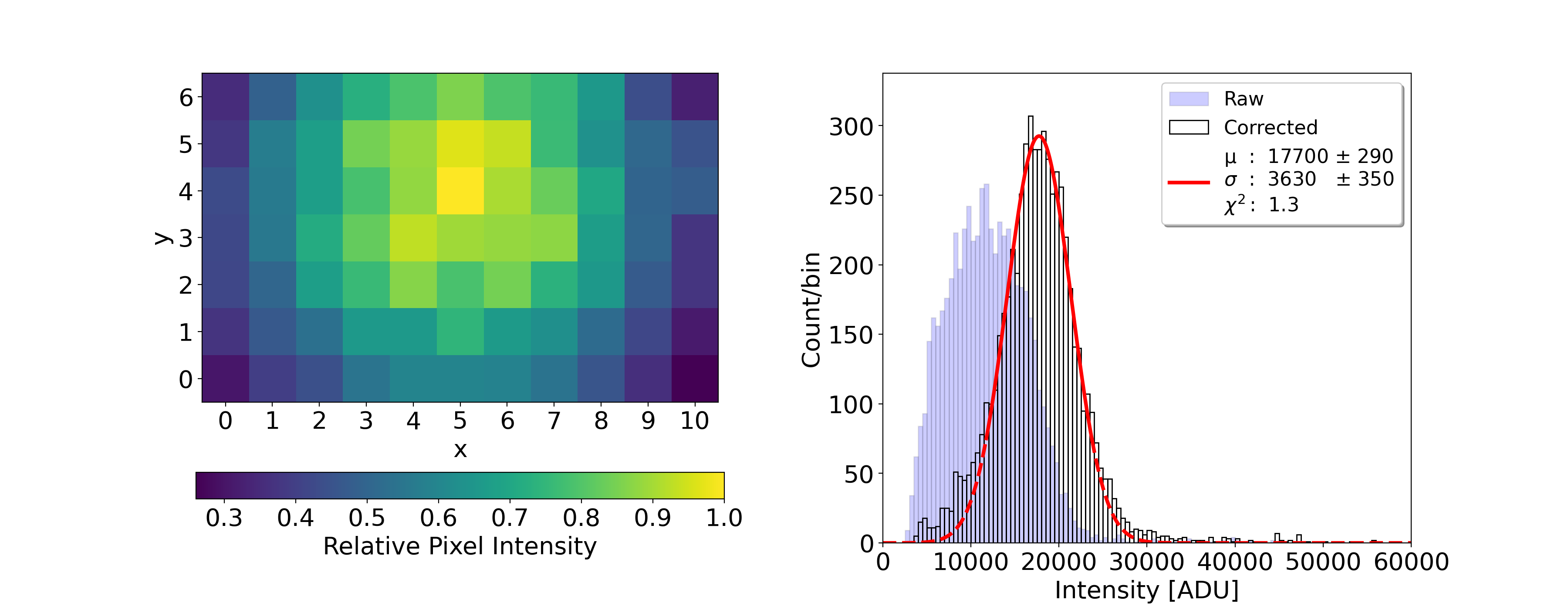}
\caption{Left: Effective intensity map generated from $^{55}$Fe x-rays. Right: $^{55}$Fe spectrum before (pale blue bars) and after (black bars with fit) applying the vignetting correction map.\label{fig:vignetting}}
\end{figure*}
\section{Vignetting corrections and energy calibrations}
\label{sec:A1}

In the presence of monoenergetic events, observed intensities in the optical readout diminish due to vignetting as we move radially away from the center of the optical axis. While K-$\alpha$ and K-$\beta$ modes both contribute to the $^{55}$Fe x-ray spectra, the \SI{5.9}{keV} K-$\alpha$ mode contributes to nearly $90\%$ of detectable x-ray emissions, so we treat the spectrum as having a monoenergetic \SI{5.9}{keV} peak and log the intensities of $^{55}$Fe x-rays in regions across the optical readout to generate an effective intensity map across the camera field. We use this map to flatfield the intensity spectrum across the readout, which is what we call vignetting corrections.

To create such a map, we discretize the camera plane into an $11\times 7$ grid of logical pixels and perform the following procedure at the end of each $^{55}$Fe run:
\begin{enumerate}
\item Select only fiducial $^{55}$Fe tracks.
\item Let $P_{ij}:\text{  }i\in[0,10]\text{  },j\in[0,6]$ represent a logical pixel in our $11\times 7$ grid. For each $^{55}$Fe track, we count the number of pixels in the track falling inside each logical pixel. We assign the track to the logical pixel, $P_{mn}$ that contains more of the $^{55}$Fe track's pixels than any other $P_{ij}$.  
\item For each logical pixel, compute the mean intensity, $I_{ij}$ of the subset of $^{55}$Fe tracks that are \textit{fully} contained in $P_{ij}$. Normalize to the logical pixel with the highest mean intensity.
\end{enumerate}

The left panel of Fig.$\,$\ref{fig:vignetting} shows the effective intensity map generated using this procedure on an $^{55}$Fe run from Science Run 1. If we normalize the intensity of the logical pixel with the highest mean intensity to unity, then, for a given $^{55}$Fe track with raw intensity $I$, that was assigned to logical pixel $P_{mn}$ (with relative intensity $I_{mn}$), its vignetting-corrected intensity $I_\mathrm{cor}$ is given by
\begin{align}
\label{eq:Vignetting_correction}
I_\mathrm{cor} = \frac{1}{I_{mn}}I.
\end{align}
The black outlined histogram in the right panel of Fig.$\,$\ref{fig:vignetting} shows $I_\mathrm{cor}$ for each fiducial $^{55}$Fe track in the run. Contrast this to the raw intensity spectrum shown in pale blue. We calibrate energy by assigning \SI{5.9}{keV} to the value of $I_\mathrm{cor}$ corresponding to the peak of the single-peak Gaussian fit to the $I_\mathrm{cor}$ spectrum.

We generated effective intensity maps for each $^{55}$Fe run recorded during Science Runs 1 and 2 and performed vignetting corrections when computing the energies for all tracks during D-D runs. We found that at a given voltage across the GEMs, the $^{55}$Fe peak varies over the course of the day. Thus we need to choose which $^{55}$Fe run to use when assigning energy calibrations and vignetting corrections for D-D runs. The D-D runs analyzed in this article use calibration information (including the effective intensity map) from the nearest-in-time $^{55}$Fe calibration run that had the same voltage across the GEMs as the given D-D run. For the given D-D run with associated $^{55}$Fe calibration information, we compute the energy of each track in the D-D run using the following procedure:
\begin{enumerate}
    \item Assign to the track the logical pixel, $P_{mn}$ that contains more of the track's pixels than any other logical pixel.
    \item Use Eq.$\,$(\ref{eq:Vignetting_correction}) to compute $I_\mathrm{cor}$.
    \item Let $I_\mathrm{cor}'$ be the vignetting-corrected intensity of the peak of the $^{55}$Fe spectrum of the calibration run associated with the current D-D run. The energy of the track is computed as
    \begin{align}
    E = \frac{\SI{5.9}{keV}}{I_\mathrm{cor}'}I_\mathrm{cor}.
    \end{align}
\end{enumerate}

We acknowledge that the following improvements could be made to improve energy estimates during D-D runs
\begin{enumerate}
\item Effective gain tends to decay with increasing D-D exposure at a given voltage across the GEMs. This means that interpolating the $^{55}$Fe peak intensities versus time to assign $^{55}$Fe peak intensities at all times during a D-D run would be advantageous compared to our current approach of assigning the nearest-in-time $^{55}$Fe calibration performed with the same voltage across the GEMs as the D-D run.
\item Rather than assigning logical pixels on a track-by-track basis to create effective intensity maps, we could instead assign a logical pixel to each pixel in a track and perform vignetting corrections for each individual pixel. This change may substantially improve energy estimates for long tracks (especially alphas and protons) that span many logical pixels.
\end{enumerate}
While our current energy calibration procedure provides sufficient energy resolutions for resolving the fluorine and carbon endpoints from \SI{2.5}{MeV} neutrons produced from the D-D-generator (Fig.$\,$\ref{fig:realtime}), these further improvements to our calibrations may sharpen these endpoints in the NR energy spectrum.

\bibliography{apssamp}

\end{document}